%% file: main.tex
\documentclass[12pt,a4paper]{article}
\usepackage{authblk}
\usepackage{soul, ulem}
\usepackage{verbatim, booktabs, multirow, hyperref,dirtytalk}
\usepackage{graphicx, rotating}
\usepackage[round]{natbib}
\usepackage[utf8]{inputenc}
\usepackage{enumerate}
\usepackage{euscript}
\usepackage{amsmath, amsthm,amsfonts, amssymb}
\usepackage[a4paper]{geometry}
\usepackage{tikz}
\usetikzlibrary{calc}
\hypersetup{colorlinks,linkcolor={blue},citecolor={blue},urlcolor={red}}
\usetikzlibrary{positioning}
\usetikzlibrary{arrows.meta}
\usepackage{caption}
\usepackage{subcaption}

\usepackage{pgfplots}
\usepgfplotslibrary{dateplot}
\usepackage{pdfpages}

\usepackage{float}

\newtheorem*{example*}{Example}

\theoremstyle{definition}
\newtheorem{hyp}{Hypothesis}

\newcommand{\red}{\textcolor{red} }

\usepackage{setspace}

\usepackage{verbatim, booktabs, multirow, hyperref}
\usepackage{bbm}
\usepackage{lmodern}
\usetikzlibrary{intersections}

\usepackage{tikz}
\usetikzlibrary{positioning}
\usepackage{tikz-3dplot}
\usetikzlibrary{arrows, patterns, decorations.pathmorphing,backgrounds,positioning,fit,petri,trees}
\usepackage{pgfplots}
\usepackage{pgfplotstable}

\usetikzlibrary{arrows}
\usetikzlibrary{calc}
\usetikzlibrary{positioning}
\usetikzlibrary{arrows.meta}
\usetikzlibrary{decorations.pathreplacing,calligraphy}
\usetikzlibrary{shapes.geometric}

\definecolor{FUMpurple}{RGB}{115,0,173}
\definecolor{MainColor}{HTML}{a75477}
\definecolor{SecondaryColor}{HTML}{5a53a7}
\definecolor{BrickRed}{HTML}{cb4154}

\usepackage{pifont}
\newcommand{\cmark}{\text{\ding{51}}}%
\newcommand{\xmark}{\text{\ding{55}}}%
\pgfplotsset{compat=1.16}

\begin{document}

\title{Motives for Delegating Financial Decisions\thanks{We are indebted to Maximilian Germann, Marco Lambrecht, Lukas Mertes, J\"org Oechssler, Christoph Siemroth, Matthias Stefan  and audiences at the Southern Europe Experimental Team's Meeting in Valencia and the meeting of the Society for Experimental Finance in Sofia for helpful comments and suggestions.  This work was supported by the Economic and Social Research Council [grant number ES/T015357/1].
The experimental protocol was approved by the Ethics Committee of the University of Essex, ETH2122-0711. For the purpose of Open Access, the authors have applied a CC BY public copyright licence to any Author Accepted Manuscript (AAM) version arising from this submission.}}
\author{Mikhail Freer\thanks{University of Essex, email: m.freer@essex.ac.uk} \and Daniel Friedman\thanks{University of Essex and University of California Santa Cruz, email: dan@ucsc.edu}    \and  Simon Weidenholzer\thanks{University of Essex, email: sweide@essex.ac.uk}
}
\date{14 April 2024}
\maketitle

\begin{abstract}
Why do some investors delegate financial decisions to supposed experts? We report a laboratory experiment designed to disentangle four possible motives. 
Almost 600 investors drawn from the Prolific subject pool choose whether or not to delegate a real-stakes choice among lotteries to a previous investor (an ``expert'') after seeing information on the performance of several available experts. 
We find that a surprisingly large fraction of investors delegate even trivial choice tasks, 
suggesting a major role for the blame shifting motive. 
A larger fraction of investors delegate our more complex tasks, suggesting that decision costs play a role for some investors. Some investors who delegate choose a low quality expert with high earnings, suggesting a role for chasing past performance. We find no evidence for a fourth possible motive, that delegation makes risk more acceptable. 
\end{abstract}

\vspace{.5cm}



\thispagestyle{empty}

\setstretch{1.4}

\newpage
\setcounter{page}{1}

\section{Introduction}
Traditionally the delegation of financial decisions incurred substantial fees and was used only by a few wealthy investors. However, the recent rise of trading platforms such as {\it eToro} and {\it ZuluTrade} encourages a much larger set of investors to delegate their financial decisions by designating another investor (referred to below as an ``expert,'' without prejudging actual expertise) whose trades will be copied.
See e.g.\ \cite{pelster2018fear} and \cite{apesteguia2020copy} for discussion of the scope and operational details of such platforms.

Such recent innovations make it imperative to understand better \textit{why} investors delegate and \textit{which experts} they choose.
For example, investors may gain substantially if they delegate at low cost to high quality experts, but can be harmed if they chase the past performance of recently lucky but low quality experts,  or if they choose an expert who maintains an excessively risky portfolio. 

In this paper we seek to disentangle possible motives for delegation, and to investigate investors' responsiveness to delegation cost. 
It is quite difficult to do so using field data, which do not permit us to observe investors' and experts' characteristics such as risk preferences, and have little variation in key variables such as delegation costs, the nature of the delegated task or the available information on experts.
Therefore we conduct a human subject laboratory experiment that controls and systematically varies those key variables.   

As elaborated below in our survey of previous literature, we identify four possible motives for delegation: 
(i) chasing past performance; 
(ii) blame shifting, or more generally, aversion to personally making decisions;
(iii) reducing decision costs, which may be excessive for less knowledgeable investors facing complex decisions; and
(iv) increasing risk tolerance when delegating, either because delegation reduces the investor's subjective utility cost for risk or else because it reduces her perception of the riskiness of the expert's choice.  

These motives differ sharply in their implications for investor welfare. Chasing past performance is a belief distortion that tends to reduce investors' wealth, and blame shifting will incur unproductive costs when there is a fee associated with delegation. 
On the other hand, 
reducing decision costs may be rational and beneficial for an investor, and (to the extent that it is consistent with investor well-being) increasing risk tolerance may also be beneficial. 

In our experiment, human subjects make a series of real-stakes investment decisions, and have the option to  delegate some of those decisions to an expert. 
We pre-selected a small panel of experts from participants in preliminary sessions that included all the choice tasks faced by participants in the experiment. 
Delegation  means that the investor is assigned the lottery chosen by the selected expert. 
Of course, the realized lottery outcome may differ from the outcome that the expert experienced in the preliminary session.

The experiment varies the information available when an investor selects an expert. In the AllInfo treatment, investors see each expert's realized final payoff (a very noisy signal of expertise), quality rating (a less noisy signal) and risk rating (based on the expert's revealed risk preferences). In sessions with the NoPay treatment, investors see the other two expert characteristics but not realized payoff, and in the NoQual treatment they see the experts' characteristics except the quality rating.
We also vary the cost of delegating across sessions; it is either low (10 pence, about 1.5\% of average earnings) or high (100 pence, about 15\%). 

Each investor faces three different levels of investment task complexity. A {\it trivial} task is choice between two binary lotteries with identical expected payoff and variance.
A {\it simple} task is choice among several lotteries that differ in expected payoff and variance, some of which are dominated by others.
Finally, a {\it complex} task in our experiment is portfolio allocation over the set of simple lotteries. 
Thus, while success in the trivial task is purely driven by luck, there is a role for skill in the simple task 
and even more so in the complex task. Moreover, while risk preferences are irrelevant for choice in the trivial task, they are important for choice in the simple and complex tasks.

Those treatments enable us to identify the different motives for delegation: we compare across treatments the fraction of investors who choose to delegate and compare the attributes of the selected experts when investors choose to delegate.
In the trivial task, for example, reducing decision costs and increasing risk tolerance can play no role. Hence the surprisingly high frequency of delegation we observe in the trivial task with NoPay (thus eliminating chasing past performance) is evidence of blame shifting.

We also find evidence of decision costs, in that delegation is significantly more frequent in the complex task than in the simple and trivial tasks. Evidence for chasing past performance comes from the disproportionate tendency of investors who delegate to choose an expert with a high realized final payoff. We see little evidence of the risk tolerance motive.

In the next section we summarize previous literature, with a focus on the possible motives for delegation and how they can be identified.
The third section lays out our laboratory procedures and experiment design, and spells out the hypotheses on delegation motives that we will test. Results are collected in the following section, beginning with an overview using descriptive summary statistics. Later subsections present inferential statistics to identify the strength and significance of the various motives for delegation, and to analyze which characteristics matter for the selection of experts. 
We were surprised to discover that the decision to delegate is quite sensitive to conflicts among reported characteristics of experts. 
We also were surprised by some null results, e.g., own risk preferences and decision quality play no detectable role in the delegation decision.
In the final section 
we discuss such matters 
and suggest directions for future research. Online Appendices collect supplementary data analysis, technical details, and instructions to subjects.

\section{Literature}
We first discuss literature that is closest to the present paper, and then review literature that proposes motives for delegation.

\subsection{Nearest neighbors}

Our work is closely related to recent work on delegation in financial decision making. \cite{apesteguia2020copy}, like the present paper, report an experiment where investors may decide to delegate financial decisions  to their peers. They show that a substantial fraction of investors does so by either directly copying previously successful investors by the click of a button or manually implementing investment strategies which are similar to those of the most successful peers. Since success is mainly driven by luck and since investors who previously took on a lot of risk appear on top of the earning rankings, \cite{apesteguia2020copy} find that copy trading may lead to a substantial increase in risk taking. 

The present study extends the design of \cite{apesteguia2020copy} by varying the complexity of the underlying task and the information investors receive about the experts. When our investors do not have access to information on experts' decision quality, we confirm that a substantial fraction of subjects chooses to delegate to experts with previously high earnings.\footnote{Other studies besides \cite{apesteguia2020copy} finding an important role for previous earnings in the choice of ``experts'' include \cite{huck1999learning}, \cite{offerman2002imitation}, \cite{apesteguia2010imitation} and \cite{huber2010hot}.}
However, we find that 
the share of investors choosing high earning experts substantially decreases when earnings information is augmented by information on decision quality.  

\cite{holzmeister2022delegation} study delegation in a more traditional setting using a subject pool of finance professionals and members of the general public, both drawn from the Swedish population. Each investor is matched with an expert who is either a human with fixed or aligned incentives or a robo-advisor, i.e., an appropriately programmed algorithm.\footnote{\cite{germann2023algorithm}, \cite{gaudeul2023trade} and \cite{lambrecht2023benefits} also report experiments where investors can delegate to an algorithm.}
In contrast to our work, they offer no information on earnings or decision quality, nor do their investors choose among experts. 
Their investors  indicate i) whether they want to delegate their decision to their expert, ii) their maximum willingness to pay for delegation and iii) how much risk they want the expert to take on their behalf.  \cite{holzmeister2022delegation} find that investors delegate most frequently to the algorithm, and least frequently to experts with fixed preferences. Moreover, delegation is positively correlated with general trust and blame shifting tendencies as elicited in a survey. (Our experiment uses observed behavior to confirm the importance of blame shifting.) 
Moreover, \cite{holzmeister2022delegation} find that investors ask the expert to take more risk than they believe themselves to have taken, 
consistent with the increasing risk tolerance motive.\footnote{{Analysing the same data-set as \cite{holzmeister2022delegation}, \cite{stefan2022you} identify a significant problem with risk communication in the sense that while finance professionals in general take into account the client's desired risk level,  the constructed portfolios show considerably overlap across the different requested risk levels.}} 
By contrast, we find that the fraction of delegation 
does not increase as we move from a task where there is no scope for risk communication to one where risk communication matters. This suggests that while the increasing risk tolerance motive may matter in the decision of how much risk investors ask the expert to take, that motive does not appear to explain why investors choose to delegate in the first place. 

\subsection{Motives for Delegation}

Existing literature suggests four possible motives for delegating financial decisions.
Firstly, under {\it chasing past performance} \citep[see e.g.][]{sirri1998costly}, investors believe that by delegating they will be able to realize high profits, as the expert did in the past.
That is, the choice is rational given possibly distorted beliefs.\footnote{
E.g., a fallacious belief in an expert's  ``hot streak'' as in \cite{huber2010hot}.
More generally, chasing past performance in financial decisions is a form of success-based imitation.
Of course, depending on the link between an action's current earnings and future earnings, such imitation may actually decrease payoffs \citep[see e.g.][]{vega1997evolution,huck1999learning,offerman2002imitation} as well as possibly increase them  \citep[see e.g.][]{schlag1998imitate,apesteguia2018imitation}. }
However, in environments where success is purely driven by luck --- so skill plays no role ---  chasing past performance may be detrimental to  welfare \citep[see e.g.][]{offerman2009imitation}. Note that chasing past performance as a motive for delegation  can be eliminated by blocking information on experts’ previous payoff success. 
Conversely, when previous payoff information is available, the fraction of investors delegating to experts with high earnings compared to the fractions for experts with other desirable characteristics can help isolate the chasing past performance motive. 

Secondly, investors may delegate to {\it save on decision costs} or to make better decisions in a complex environment; see \citealt{conlisk1980costly,pingle1996modes} and the survey of \citealt{alos2009imitation}.
That is, investors believe that the environment is too complex for them to make optimal decisions, except perhaps by incurring excessively large subjective decision costs.\footnote{
Similarly, individuals may choose to delegate to save on information-gathering costs behind the observed choices as in \cite{sinclair1990economics}.}
Such investors may be rational in the usual neoclassical sense, but recognize their own cognitive limitations. If true expertise is available at moderate cost, it is rational for them to hire it. 
Thus, while the decision cost motive allows no role for delegation in transparently simple tasks or those purely driven by luck,  it offers an increasing role for delegation in cognitively more complex environments. Also, the  decision cost motive should cause investors who delegate to choose high quality experts.\footnote{
{The importance of the decision cost motive is documented in \cite{mertes2023information} who present an experiment where subjects may adjust their own actions after receiving objective information on their peers' performance. In this setting peers with higher expertise are more likely to be followed and subjects who have low confidence in their own decisions are more inclined to follow. }
}

Thirdly, under the {\it blame shifting} motive, agents delegate because they want somebody to blame in case things don’t work out; see for example \cite{bartling2012shifting} or \cite{gurdal2013blame}. Blame shifting may be seen as an form of \textit{decision avoidance} \citep{anderson2003psychology} where individuals shy away from decisions to \textit{avoid regret}. Shifting blame to a delegatee may also allow investors to resolve \textit{cognitive dissonance};
as  \citet{chang2016looking} emphasizes, realizing losses based on personal decisions forces investors to admit that their own past choices were mistakes.
The literature suggests that the blame shifting motive is as present in simple and pure luck tasks as it is in complex tasks. 
Nor does the blame shifting motive per se suggest a role for experts' characteristics in the selection of an expert when delegating.  

Finally,  according to the {\it increasing risk tolerance} motivation \citep[see e.g.][]{gennaioli2015money}, trust in the expert may reduce an investor’s  utility cost of taking risk. Thus, delegating to experts may encourage the investor to take on greater risk.\footnote{The importance
of trust is echoed by \cite{loos2019trust} who provide experimental evidence that the higher the level of trust in a given advisor is, the more risk clients ask this advisor to take.} 
Foerster et al. (2018) confirm empirically that higher trust
in advisers results in higher risk-taking by investors.
This motive has no observable implications in settings where all actions have the same degree of risk. In other settings, it implies that investors will tend to choose experts whose risk tolerance equals or exceeds their own. 

\section{Design}
\begin{figure}[ht]
\centering
\resizebox{1\linewidth}{!}{
\begin{tikzpicture}
\node[draw,
    rectangle,
] (t1) at (0,0) {Investment};
\draw[] (0,-1.5) node[above] {$t=1$};

\node[draw,
    rectangle,
] (t2) at (3,0) {Investment};
\draw[] (3,-1.5) node[above] {$t=2$};

\node[draw,
    rectangle,
] (t3) at (6,0) {Investment};
\draw[] (6,-1.5) node[above] {$t=3$};

\node[draw,
    rectangle,
] (t4) at (9,0) {Investment};
\draw[] (9,-1.5) node[above] {$t=4$};

\node[draw,
    diamond,
    aspect=2,
] (t5) at (13,0) {\bf Delegate?};
\draw[] (13,-1.5) node[above] {$t=5$};

\node[draw,
    rectangle,
] (No) at (13,2) {Investment};

\node[draw,
    rectangle,
] (Yes) at (17,0) {Next block};

\draw[thick, ->] (t1) -> (t2);
\draw[thick, ->] (t2) -> (t3);
\draw[thick, ->] (t3) -> (t4);
\draw[thick, ->] (t4) -> (t5);

\draw[thick, ->] (t5) -> (Yes) node[above, midway] {\bf Yes};
\draw[thick, ->] (t5) -> (No) node[left, midway] {\bf No};
\draw[thick, ->] (No) -- ++(4,0) --  (Yes);

\end{tikzpicture}
}
\caption{
Timeline for each block. The first 4
\textit{Investment} decisions must be made by the investor. The fifth decision can be delegated to a chosen expert.
}
\label{fig:BlockTimeline}
\end{figure}
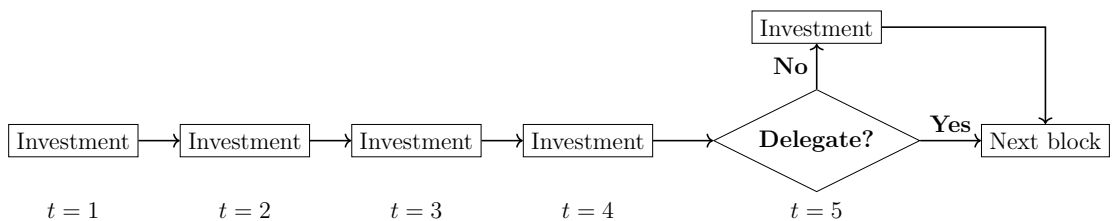

The experiment involves a sequence of individual decision-making tasks.
The sequence consists of three blocks, each proceeding as in Figure \ref{fig:BlockTimeline}.
Our human subject investors start each block with four different investment decisions.
The block concludes with a fifth and final decision that can be delegated to a chosen expert or, if the investor prefers, can instead be taken by the investor herself.
The level of complexity is constant within each block but differs across the three blocks as detailed below.

At the end of the experiment, one of the blocks is randomly selected for payment. For that block,
one out of the first four investment decisions is randomly selected for payment, and the fifth decision is always also selected. 
Thus the fifth (delegation) decision is for higher stakes because its realized (state dependent) outcome will be paid with
probability $\frac{1}{3}$, versus payment probability
$\frac{1}{12}$ for the other four decisions in the block.
There is no feedback until the end of the experiment where
investors observe the resolution of uncertainty only for the two rounds of the randomly selected block. This design choice rules out path dependency by making sure that observations do not depend on  previous idiosyncratic realizations of payoffs.

\subsection{Investment Decision}
All investment decisions involve lotteries with two equally likely states of the world, denoted Heads and Tails. 
Menus of  lotteries have three levels of \textit{complexity}, as illustrated in
Figure \ref{fig:InvestmentDecision}; see
Appendix \ref{appendix:Instructions} for the complete set of menus.

\begin{figure}[hp]
    \centering
\begin{subfigure}[h]{.8\linewidth}
    \caption{Trivial Decision Task}
    \includegraphics[width = 1\linewidth]{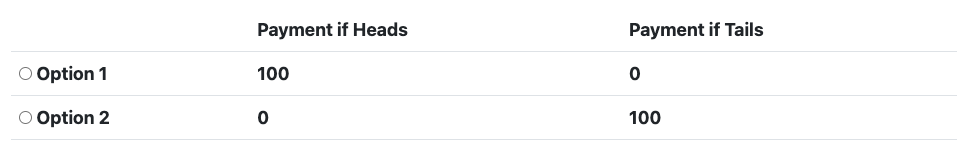}
    \label{fig:TrivialTask}
\end{subfigure}

\vspace{.5cm}

\begin{subfigure}[h]{.8\linewidth}
    \caption{Simple Decision Task}
    \includegraphics[width = 1\linewidth]{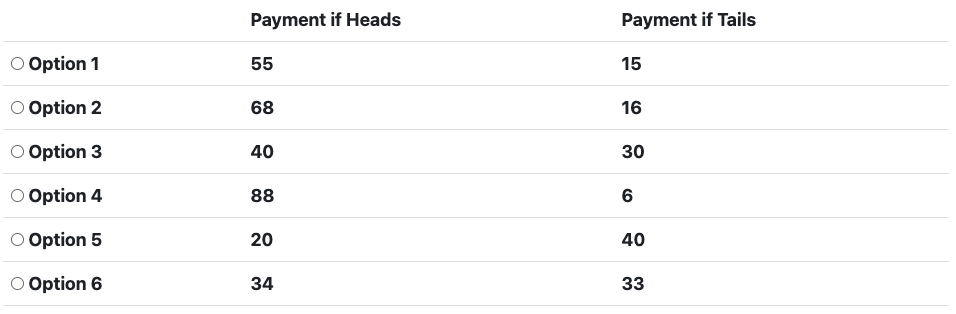}
    \label{fig:SimpleTask}
\end{subfigure}

\vspace{.5cm}

\begin{subfigure}[h]{.8\linewidth}
    \caption{Complex Decision Task}
    \includegraphics[width = 1\linewidth]{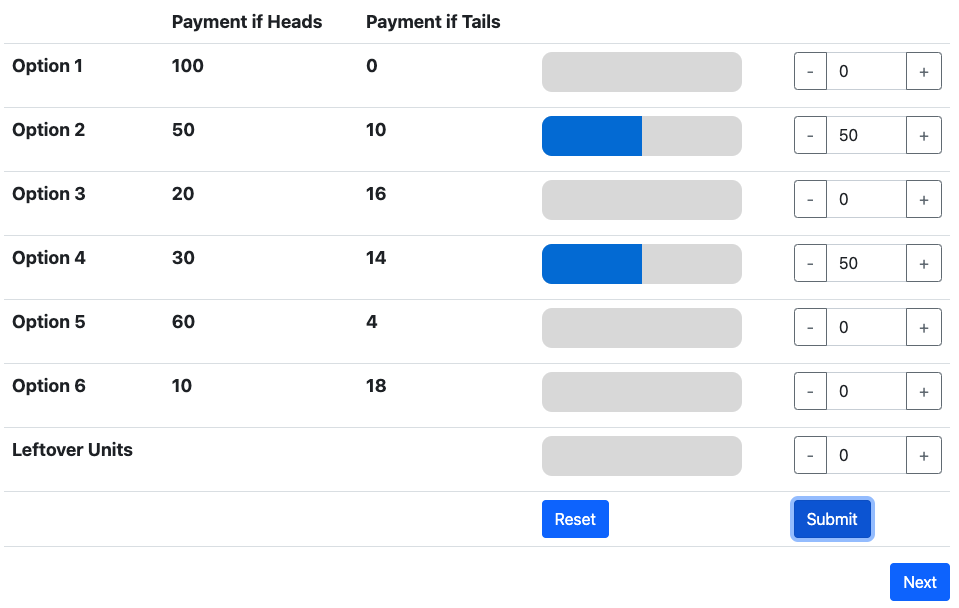}
    \label{fig:ComplexTask}
\end{subfigure}

\caption{Investment Decision. Investor clicks a radio button for tasks (a) and (b), and clicks or drags to adjust portfolio weights $w_i$ in task (c). The blue Submit button appears only when Leftover Units  $=100-\sum_{i=1}^6 w_i =0$.}
\label{fig:InvestmentDecision}
\end{figure}

Investors in a \textit{trivial task}, exemplified in Figure \ref{fig:TrivialTask}, choose  between a given lottery and an identical lottery that switches Heads and Tails.
Here the investor's choice has no impact on the ex-ante payoff distribution, and the realized earnings depend only on blind luck.
A \textit{simple task} as illustrated in Figure \ref{fig:SimpleTask} is
similar to popular risk-preference elicitation tasks involving choice from a finite menu of lotteries
\cite[e.g.,][]{binswanger1980attitudes,eckel2002sex}.
Here our investors have six options, four of which are optimal for some range of risk preferences.
Each of the  other two options (lines 1 and 5 in the Figure) is first-order stochastically dominated by another option, since Heads and Tails are equally likely.
The investor's choice therefore can either reveal a range for her risk preferences or 
else can reveal poor decision quality. 

In a \textit{complex task} as illustrated in Figure \ref{fig:ComplexTask},
investors choose a portfolio of risky assets; the user interface is a variant on the ``budget jars'' interface of \cite{friedman2022varieties}. The menu of pure assets is the same as for the simple task, but here investors use sliders to pick a particular portfolio from the five-dimensional simplex of mixtures.
This task is cognitively complex in that it is high dimensional and makes it easy to select a dominated option. 
The user interface is also mechanically more complex in that it uses sliders rather than radio buttons.


For each subject, we independently randomize the presentation order of lotteries within a task menu, as well as the sequence of five tasks within a block, and the sequence of three blocks. 


\vspace{-.1in}
\subsection{Delegation Decision}
\begin{figure}[htb]
    \centering
    \includegraphics[width = 1\linewidth]{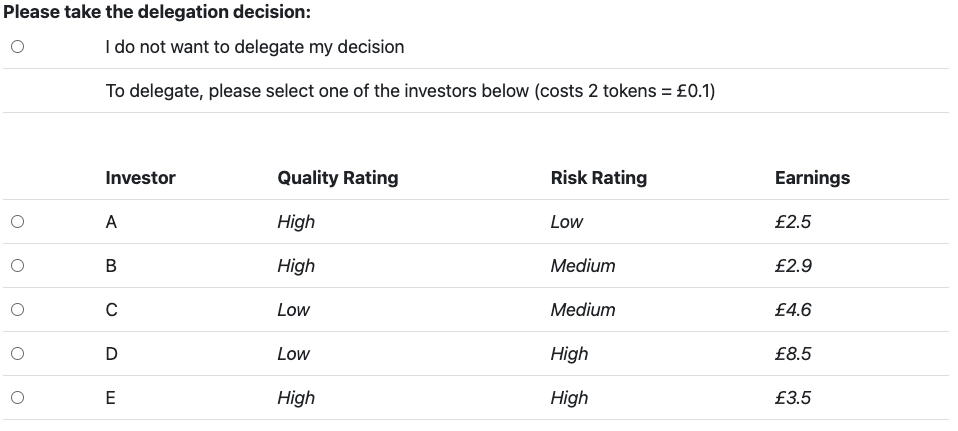}
    \caption{Delegation Decision Task}
    \label{fig:DelegationTask}
\end{figure}

\vspace{-.1in}
As noted earlier, the fifth and final task in each block 
has higher probability of being chosen for payment and allows the investor to delegate. 
Here the investor either opts to make the  choice on her own, or else selects one expert from the given panel of five experts.
In the latter case, the selected expert's choice from that menu will then be employed, although the realization (Heads or Tails) and the resulting payment may be different. 
See Figure \ref{fig:DelegationTask} for the user interface.

The panel of five experts is constructed as follows. In preliminary sessions, 60 subjects completed three blocks of 5 tasks each, exactly like the main sessions just described except that the fifth task was a direct investment task no different from the other four.

From the choices in the simple and complex blocks of each of those 60 potential experts, we construct a \textit{quality rating} using the \citet{afriat1973system} Critical Cost Efficiency Index, as detailed in Online Appendix \ref{App:CCEI}. 
Roughly speaking, the
CCEI is ``money left on the table,'' measured as the minimal share of the wealth taken from a subject that would make their choices consistent with maximizing the expectation of some utility function.
We assign a High quality rating to those in the lowest CCEI quartile among the 60 potential experts and a Low quality rating to those in the top CCEI quartile.
We also assign a \textit{risk rating} to each potential expert based on their average coefficient of relative risk aversion implied by each non-dominated choice in the simple and complex blocks.  A potential expert is rated as High, Medium or Low risk tolerance according to whether that average is in the lowest, middle or upper third of the distribution across the 60 potential experts.

In addition, we note the realized \textit{payment} of each potential expert.
As we point out in the instructions to subjects,\footnote{
The instructions read, 
``Their final payments, like yours, come from the last round and one of the other four rounds in a single randomly chosen block. Please note that it is often from a different block (and round) than that for which you delegate.''
} payment is a very noisy signal of experts' skill. 
For example, with probability $1/3$, the earnings depend on the expert's choice in a trivial task where skill is irrelevant, and earnings in the other tasks depend sensitively on the riskiness of the expert's choice and/or on the realized state (H or T).
Hence, objectively speaking, earnings add little if any useful information about experts when 
quality and risk ratings are known.

From among those 60 subjects, we selected 5 to designate as experts in the main sessions.
We did not include anyone dominated by any of the other four in terms of earnings and risk and quality ratings. Three of the selected experts had High quality rating and different Risk ratings (L, M, H).
The other two selected experts had Low quality rating and Medium or High risk tolerance.
Given our elimination of dominated subjects, it is not surprising to see that the two Low quality experts have the highest realized earnings.
(We will later see that that contrast is helpful for investigating the \say{chasing past performance} motive.)
To those five experts we randomly assign the labels A, B, C, D, and E  when presenting them to an investor.

Investors are aware that the experts are other human subjects recruited on the same platform who previously completed the same set of investment tasks.
To avoid implicit endorsement, the instructions and interface refer to experts only as ``previous investors.''
Investors understand that delegation means that their choice will be the chosen expert's
choice on that task, but their payoff is typically different because the randomly selected task is usually different and the realized state may also be different.

Delegation is costly.
Each investor is endowed with \pounds1 at the beginning of the experiment.
If an investor chooses to delegate their decision in a block chosen for payment, then they pay the cost, which (to avoid the possibility of negative earnings) does not exceed \pounds1.

\subsection{Between-Subject Treatments}
As mentioned above, each subject faced three blocks, one for each level of the complexity treatment.
Between subjects we varied two other treatments: delegation cost and information on experts.

\paragraph{Delegation cost.}
In the \textit{Low} cost condition, the subject has to pay \pounds 0.1 (10 pence) to delegate the decision.
This small positive cost will ensure that a neo-classically rational  subject would not delegate their decision in the trivial investment task. 
In the \textit{High} cost condition, the subject has to pay \pounds 1 to delegate. 
Of course, this cost is deducted from realized earnings only when the investor delegates in the  block randomly selected for payment.

\paragraph{Information on experts.}
In the \textit{AllInfo} condition, subjects see all three characteristics for each expert on the panel: risk rating, quality rating, and earnings.
In the \textit{NoPay} condition, subjects see only the risk and quality ratings of the experts but not their earnings.
In the \textit{NoQual} condition, subjects see experts' risk rating and earnings but not their quality rating.

Thus the experiment uses a $3\times 3 \times 2$ factorial design, with three complexity conditions within subjects,
three information conditions between subjects, and two delegation cost conditions between subjects.

\subsection{Experimental Procedures}
All experiments were conducted online using the Prolific platform.
The sample consists of UK residents no older than 50 who are native English speakers.
The experimental software is coded using oTree \citep[][]{chen2016otree}.
The experiments were conducted from September 2022 to February 2023.
Preliminary sessions had 60 participants (no delegation). We had 
589 participants with the delegation option in the experiment analyzed in the next section,
with between 97 and 99 participants in each of the 6 between-subject treatments.
The experiment lasted on average 15 minutes and subjects' earnings are about \pounds7 on average.

\subsection{Testable Hypotheses}

Our design leads to natural tests of the various motives for delegation discussed in the literature review section, as follows.

\begin{hyp}[\textbf{Chasing Past Performance}]
\label{h:CPP}
(a) The delegation rate will be significantly lower in the treatment NoPay where experts' payoffs are not observed. 
(b) In treatments where experts' payoffs are observed, delegation will be disproportionately to those with highest payoffs.
\end{hyp}

\begin{hyp}[\textbf{Blame Shifting}]
\label{h:BlameShifting}
The delegation rate will be substantially greater than zero even in the Trivial task, in all information conditions.
\end{hyp}

\begin{hyp}[\textbf{Reducing Decision Costs}]
\label{h:DecisionCosts}
The delegation rate will be (a) higher in  Simple tasks than in Trivial tasks and
(b) considerably higher in Complex tasks than in Simple tasks.
(c) In treatments where experts' quality is observed, delegation will be disproportionately to those with high quality.    
\end{hyp}

\begin{hyp}[\textbf{Increased Risk Tolerance}]
\label{h:RiskTolerance}
(a) The delegation rate will be considerably higher in simple tasks than in trivial tasks, but not appreciably higher in complex than simple tasks.
(b) In simple and complex tasks, delegation will disproportionately be to experts whose risk rating exceeds the investor's.    
\end{hyp}

Hypotheses \ref{h:CPP} and \ref{h:BlameShifting} follow directly from the discussion in the literature review,
bearing in mind that, in testing \autoref{h:BlameShifting}, we use the term ``blame shifting'' in a broad sense  to include other forms of decision avoidance.
Nuances of \autoref{h:DecisionCosts} and \autoref{h:RiskTolerance}, however, may deserve additional comment. 
As the names suggest, the Complex task  seems much more cognitively  (and mechanically) demanding than the other two tasks, so parts (a) and (b) of the decision costs \autoref{h:DecisionCosts} suggest that the contrast between Complex and Simple could be more dramatic than the contrast between Simple and Trivial. On the other hand, risk attitudes play a role only in the Complex and Simple tasks. (If anything, we'd expect risk to be at least as salient in Simple as in Complex, since the latter offers diversification opportunities.)
Hence \autoref{h:RiskTolerance}a suggests a less dramatic contrast (with ambiguous sign) between Simple and Complex than between Trivial and the other two tasks.

Although not directly relevant to our main concern, distinguishing among motives for delegation, it is also natural for economists to examine price elasticity.  
Therefore we will also test the following.

\begin{hyp}[\textbf{Price Elasticity}]
\label{h:Price}
Delegation rates will be lower in the High Cost treatment.     
\end{hyp}



\section{Results}
\label{sec:Results}


\input{paper_Figs_and_Tabs/treatment_effects}

Before presenting the evidence pertaining to our main questions, we first offer an overview of the data and treatment effects.
Figure \ref{fig:complexRes} pools across cost and information treatments, and shows that there is  a high delegation rate in each complexity condition. 
About half of our investors delegate even the trivial and simple choice tasks. About 60\% delegate the complex task; later we will see that this increase is statistically significant.
Figure \ref{fig:costbycomplexRes} pools across complexity and shows that delegation is insensitive to cost in the AllInfo and NoQual conditions.
We will see later that the reduction in delegation frequency is statistically significant in the NoPay (experts' realized earnings not shown) condition.


\subsection{Whether to delegate?}\label{sect:Whether}

We now take a deeper look at the delegation process.
Table \ref{tab:DelegationGrandRegressionS} reports a regression of the dummy variable for delegation (three observations per subject) on treatment variables.
This linear probability model indicates that our investors delegate in almost exactly half of the available opportunities in the baseline treatment (Trivial task, third task, AllInfo, Low cost).
The estimated effects of seeing the task sooner (in the first or second block) or withholding information on experts (NoPay or NoQual) are all negligible and statistically insignificant.
Contrary to Hypotheses \ref{h:DecisionCosts}(a) and \ref{h:RiskTolerance}(a), the effect of changing the task from Trivial to Simple is also negligible. 
Since the delegation rate is so high in the Trivial task even in the NoPay condition, we conclude that,
consistent with Hypothesis \ref{h:BlameShifting},
blame shifting
is a primary motivation for many of our investors.  

The Table also shows that changing the task to Complex increases the delegation rate by 11 percentage points.
That increase is highly significant, indicating that,
consistent with Hypothesis \ref{h:DecisionCosts}(b), complexity-related decision cost is an important motivation for some investors.
The first column of the Table indicates that increasing the cost of delegation by 90 pence decreases delegation by only an insignificant 4 percentage points in the baseline.
The second column indicates that price sensitivity is much greater in the NoPay condition, and that the decrease of $1+13=14$ percentage points is significant.
A more detailed analysis of the delegation decision in the next subsection will suggest an interpretation of this mixed evidence for Hypothesis \ref{h:Price}.


\begin{table}[htb]
\begin{center}
\begin{tabular}{l|rc@{}ll|rc@{}ll}
\toprule
& \multicolumn{4}{c|}{\bf (1)}
& \multicolumn{4}{c}{\bf (2)}                                \\ \midrule
Constant                   & $0.50$  & $\pm$ & $0.04$ & $(^{***})$ & $0.48$  & $\pm$ &  $0.05$ & $(^{***})$                \\ [.5ex]
First block                     & $0.01$  & $\pm$ & $0.03$ &            & $0.01$  & $\pm$ & $0.03$ &                           \\ [.5ex]
Second block                    & $0.00$  & $\pm$ & $0.03$ &            & $0.00$  & $\pm$ & $0.03$ &                           \\
Simple task                & $0.01$  & $\pm$ & $0.03$ &            & $0.01$  & $\pm$ & $0.03$ &                           \\ [.5ex]
Complex task               & $0.11$  & $\pm$ & $0.03$ & $(^{***})$ & $0.11$  & $\pm$ & $0.03$ & $(^{***})$                \\ [.5ex]
NoPay                      & $0.03$  & $\pm$ & $0.03$ &            & $0.10$  & $\pm$ & $0.05$ &                           \\ [.5ex]
NoQual                     & $0.04$  & $\pm$ & $0.03$ &            & $0.03$  & $\pm$ & $0.05$ &                           \\ [.5ex]
High Costs (of delegation) & $-0.04$ & $\pm$ & $0.02$ &            & $-0.01$ & $\pm$ & $0.05$ &                           \\ [.5ex]
NoPay $\times$ High Costs  &         &       &          &            & $-0.13$ & $\pm$ & $0.06$ & $(^{*})$ \\ [.5ex]
NoQual $\times$ High Costs &         &       &          &            & $0.03$  & $\pm$ & $0.06$ &                           \\ [.5ex]
\midrule
Num. obs.                  & \multicolumn{4}{c|}{$1767$}              & \multicolumn{4}{c}{$1767$}                 \\
\bottomrule
\multicolumn{7}{l}{\scriptsize{$^{***}p<0.001$; $^{**}p<0.01$; $^{*}p<0.05$}}
\end{tabular}
\caption{Linear probability model coefficient estimates ($\pm$ std error) for delegation decision. Standard errors are clustered at the individual subject level.
}
\label{tab:DelegationGrandRegressionS}
\end{center}
\end{table}

\begin{table}[htb]
\begin{center}
\begin{tabular}{l|rc@{}ll|rc@{}ll}
\toprule
& \multicolumn{4}{c|}{\bf (1)}
& \multicolumn{4}{c}{\bf (2)}                 \\ \midrule
Constant                     & $1.57$  & $\pm$ & $(0.09)$ & $(^{***})$ & $2.09$  & $\pm$ & $(0.31)$ & $(^{***})$ \\
NoPay                        & $0.29$  & $\pm$ & $(0.13)$ & $(^{*})$   & $0.31$  & $\pm$ & $(0.13)$ & $(^{*})$   \\
NoQual                       & $0.08$  & $\pm$ & $(0.13)$ &            & $0.09$  & $\pm$ & $(0.13)$ &            \\
High Costs (of delegation)   & $-0.02$ & $\pm$ & $(0.13)$ &            & $0.01$  & $\pm$ & $(0.13)$ &            \\
NoPay $\times$ High Costs    & $-0.39$ & $\pm$ & $(0.13)$ & $(^{*})$   & $-0.40$ & $\pm$ & $(0.13)$ & $(^{*})$   \\
NoQual $\times$ High Costs   & $0.08$  & $\pm$ & $(0.13)$ &            & $0.06$  & $\pm$ & $(0.13)$ &            \\
Efficiency (CCEI)           &         &       &          &            & $-0.01$ & $\pm$ & $(0.13)$ &            \\
Relative Risk Aversion &         &       &          &            & $-0.07$ & $\pm$ & $(0.13)$ &            \\ \midrule
Num. obs.                    & \multicolumn{4}{c|}{$589$}               & \multicolumn{4}{c}{$589$}   \\ \bottomrule
\multicolumn{9}{l}{\scriptsize{$^{***}p<0.001$; $^{**}p<0.01$; $^{*}p<0.05$}}
\end{tabular}
\caption{OLS regression coefficient estimates ($\pm$ standard error)
for delegation intensity.
The dependent variable is the number of times a subject delegated (0, 1, 2, or 3). }
\label{tab:DelegationIntensityS}
\end{center}
\end{table}
\vspace{-.025in}


Some nuance can be gleaned from looking at the intensity of delegation. 
Table \ref{tab:DelegationIntensityS} reports regressions of information and personal variables on the frequency of delegation (0, 1, 2 or 3 times) across our 589 investors.
The first column confirms that our investors delegate on average about half the time (1.57 of 3 opportunities). It also shows that, with low cost, the delegation frequency is 0.29 higher when investors don't see experts' realized payoff, but with high cost that increase is more than offset (-0.39);
both impacts are significant at the 5\% level.
Again, we defer interpretation until the next subsection.   

Supplementary data analysis in the Appendix shows that these results are robust to changing the baseline and to using logit instead of a linear probability model.
That analysis finds no impact of subjects' own risk preferences or decision quality on the delegation decision.
Additional subject-level variables (revealed risk preferences and decision quality, and responses to Big 5 personality questions) have almost no explanatory power for the delegation decision, nor impact on the coefficients just reported.

To summarize, we find strong evidence for
blame shifting (\autoref{h:BlameShifting}) in Tables \ref{tab:DelegationGrandRegressionS} and \ref{tab:DelegationIntensityS}; almost half of subjects delegate even trivial (and simple) tasks.
We also find significant evidence for complexity-based decision costs when we move from either the trivial or the simple task to the complex task (supporting \autoref{h:DecisionCosts}(b)) but find no significant increase when moving from trivial to simple (contrary to \autoref{h:DecisionCosts}(a)).

So far, we see no evidence for distorted beliefs that lead to chasing past performance (\autoref{h:CPP}(a)), nor for increasing risk tolerance (\autoref{h:RiskTolerance}(a)).
Delegation shows significant price sensitivity only
when investors can't see experts' realized payoffs.
Investigating how investors choose experts will shed more light on such matters.

\begin{table}[htb]
    \centering
    \begin{tabular}{l|ccccc|c}
\toprule
                          & min     & 25\%  & 50\%  & 75\%  & max   & mean  \\ \midrule
Simple $\times$ Low Costs & 0.00    &  0.04 & 0.25  & 0.48  & 1.09  & 0.30 \\
Complex $\times$ Low Costs & 0.00    & 0.05 & 0.08  & 0.11  & 0.43  & 0.09 \\
Simple $\times$ High Costs & 0.00    & 0.00 & 0.14  & 0.37  & 1.60  & 0.37 \\
Complex $\times$ High Costs & 0.00    & 0.05 & 0.07  & 0.09  & 0.39  & 0.08 \\ \bottomrule
    \end{tabular}
    \caption{Costs of Inefficiency (in GBP)}
    \label{tab:CCEIs}
\end{table}

One important question remains before turning to that investigation: is delegation potentially cost-effective?
More specifically, how does the fixed cost of delegation (10p or 100p) compare to typical amounts of money left on the table due to suboptimal decision?

Define the inefficiency cost, or money left on the table, as 1- CCEI converted to pounds.
This can be regarded as an upper bound on the usefulness of delegation, since a human expert may still leave some money on the table.
Table \ref{tab:CCEIs} summarizes the quartiles (and min and max) of observed inefficiency costs for individual subjects by treatment. Recall that inefficiency is not possible in the trivial task, and that it is more frequent (but mechanically less costly) in complex (portfolio allocation) tasks than in our simple (lottery selection) tasks.  
The Table shows that the median investor can potentially recoup the low (10p) cost of delegation for simple tasks, and almost recoup it for complex tasks. For high cost (100p), however, delegation is potentially cost-effective only for the very most inefficient investors in the simple task, and for no investor in the complex task.
We conclude that delegation is generally not cost effective except perhaps for simple tasks with low delegation costs.

\subsection{To Whom to delegate?}\label{sect:Whom}

\begin{figure}[htb]
    \centering
\begin{tikzpicture}
\begin{axis}[
    ybar,
    legend pos=north east,
    legend style={draw=none},
    legend cell align = left,
    xtick={1,3,5,7,9},
    x tick label style={align=center, text width=1.5cm},
    xticklabels={\footnotesize{A \\ 2.5HL}, \footnotesize B \\ 2.9HM, \footnotesize C \\ 3.5HH, \footnotesize D \\ 4.6LM, \footnotesize E \\ 8.5LH },
    xmin = 0, xmax = 10,
    ymin = 0, ymax = 1,
    ylabel = {Frequency}
]

\addplot[ybar,
    fill=MainColor,
    MainColor,
    bar width = .4
     ]
    coordinates {
    (1, .09)
    (3, .16)
    (5, .25)
    (7, .21)
    (9, .28)
};
\addlegendentry{AllInfo};

\addplot[ybar,
    fill=MainColor!50,
    MainColor!50,
    bar width = .4
     ]
    coordinates {
    (1, .04)
    (3, .02)
    (5, .02)
    (7, .49)
    (9, .42)
};
\addlegendentry{NoQual};

\addplot[ybar,
    fill=SecondaryColor,
    SecondaryColor,
    bar width = .4
     ]
    coordinates {
    (1, .24)
    (3, .43)
    (5, .28)
    (7, .03)
    (9, .02)
};
\addlegendentry{NoPay};
\end{axis}
\end{tikzpicture}
\caption{Relative frequency that each expert is chosen in each information condition. The ratings code in the bottom row indicates the expert's realized earnings, the quality of their decision making (Low or High), and their risk tolerance (Low, Medium, High).
E.g., expert A earned \pounds 2.5 and  exhibited High quality and Low risk tolerance.
}
\label{fig:ExpertChosenByInfo}
\end{figure}

Figure \ref{fig:ExpertChosenByInfo} shows the fraction of delegations going to each expert, i.e., their market shares,  by information condition. Experts are arbitrarily labelled A through E, with their ratings shown below the label. With AllInfo, low quality expert E with highest realized payoff (\pounds 8.5) and high risk tolerance has the largest share, about 30\%,
but even the smallest share (for expert A, whose low realized payoff seems not fully offset by high quality and low risk tolerance) is about 8\%.
When quality information is suppressed (pink bars), only the two experts with highest realized payoffs get substantial shares.
When instead payment information is suppressed (blue bars), these experts' shares almost vanish while the High quality experts split the market.

Several conclusions are suggested by Figure \ref{fig:ExpertChosenByInfo} and confirmed by a more detailed analysis in Appendix A.
Given the information they have,
our investors very seldom choose experts that are dominated.
When investors don't see experts' realized payoffs, they go for high quality experts (consistent with \autoref{h:DecisionCosts}(c)),
and they go for
high realized payoff experts when they don't see quality (consistent \autoref{h:CPP}(b)).
However, in the AllInfo treatment, there is a conflict between
experts' reported quality and payoffs.
Savvy investors realize that reported payoff is mostly random and so they rely mainly on reported quality, but many other investors rely more on reported payoffs, i.e., they chase past performance. 

The conflicting information about experts' quality and payoff itself has an impact on delegation.
Recall that in the AllInfo treatment, high quality experts have lower realized payoff than low quality experts.
We saw in the previous section that with low cost, delegation is considerably higher when the quality/payoff conflict disappears in the NoPay information condition.
Also, we saw that investors respond strongly to delegation cost only in the NoPay condition.
Perhaps that is because there investors are less distracted by the conflict and by the wide range of reported payoffs.
Experts' realized payoff varies from 250p to 850p, which might seem to dwarf even the high delegation cost of 100p, but of course that is an illusion and distraction since the variability is driven mainly by the random choice of pay period.  As we saw in the last subsection, the potential benefit of delegation is typically on the order of 10p. 
Qualitatively similar but smaller impacts are seen in the NoQual information condition.

Appendix A offers supplementary analysis { on which experts investors choose}. It shows that choice of experts is independent of task complexity (and even of the investor's own decision quality), but that with full information it is less likely when the price of delegation is high. 
Moreover,  as also documented in Appendix A, there is no correlation between the risk preferences of the investor and the risk tolerance of the expert they have chosen, 
contrary to \autoref{h:RiskTolerance}(b).


We conclude that the motive of chasing past performance is important for many of our investors, who give more weight than warranted to experts' realized payoffs.
In general, however, our investors tend to respond sensibly to changes in our information treatments, especially when the cost of delegation is high.


\subsection{Consistent Subjects}
To check the robustness of our results we now restrict attention to the subsample of  \textit{consistent} subjects, 
i.e., those who are sensitive to the complexity of the task.
A subject who delegated in the trivial task is deemed consistent if they delegate in simple and complex tasks as well.
A subject who did not delegate in the trivial task but did delegate in simple task is deemed consistent if they delegate in complex task. 
A subject who delegates in neither the trivial nor simple tasks is deemed consistent whether or not they delegate in the complex task.

\begin{table}[ht]
    \centering
    \begin{tabular}{l|ccc}
    \toprule
                    & AllInfo   & NoPay & NoQual   \\ \midrule
        Low Costs   & 72/98 (73\%)      & 59/98 (60\%)   & 49/97 (51\% ) \\
        High Costs  & 48/99 (49\%)      & 57/98 (58\%)  & 55/99 (56\%) \\ \bottomrule
    \end{tabular}
    \caption{Distribution of consistent subjects across treatments}
    \label{tab:ConsistentSubjectsTreatments}
\end{table}

\noindent
Overall, 57\% (340 of 589) of our subjects are deemed consistent. 
Table \ref{tab:ConsistentSubjectsTreatments} breaks it down across treatment conditions.
The distribution is not too different from that in overall data set.
The frequency of delegation in trivial task remains fairly high for consistent subjects, about 30\%, and it rises to 56\% in the simple task and to 81\% in complex.

As shown in on-line Appendix Table \ref{tab:FullRegressionConsistentSubjects},
regression coefficients for this restricted sample have a similar but stronger pattern than corresponding coefficients for the overall sample. 
The constant coefficient is somewhat lower (indicating a 0.32 vs 0.56 delegation rate in trivial task) 
and the incremental impacts 
of simple and complex are somewhat larger. 
This provides additional support for the conclusions
that blame-shifting is indeed a primary motive for delegation, 
and that decision cost is also an important motive.
The regression coefficients also suggest that consistent subjects are more price sensitive, 
i.e., they tend to delegate less frequently in high costs treatment than inconsistent subjects.

\section{Discussion}
Our experiment is intended to assess four different motives for delegating investment decisions to experts. To distinguish among the motives, the experiment controls for the cost of delegation and for task complexity, as well as for the available information about experts.

\begin{table}[htb]
\centering
\begin{tabular}{ll|cc}
\toprule &       & \multicolumn{2}{c}{Delegation Evidence from}   \\  \cmidrule(lr){3-4}
&       & Whether? & To Whom? \\ \midrule
\multicolumn{1}{c}{\multirow{2}{*}{Chasing Past Performance}} & H1(a) & \xmark        &              \\
\multicolumn{1}{c}{}    & H1(b) &                      & \cmark           \\ \midrule
Blame Shifting                         & H2    & \cmark               &                   \\ \midrule
\multirow{3}{*}{Reducing Decision Costs}       & H3(a) & \xmark               &                   \\
                                               & H3(b)& \cmark               &                   \\ 
                                          & H3(c)&                      & \cmark           \\ \midrule
\multirow{2}{*}{Increased Risk Tolerance}    & H4(a) & \xmark  &                   \\
                                            & H4(b) &                      & \xmark            \\ \bottomrule
Price Elasticity                           & H5    & \xmark /\cmark               &                  \\ \bottomrule
\end{tabular}
\caption{Hypothesis Test Summary. A \cmark \ (or \xmark) in the first column indicates favorable (or unfavorable) evidence presented in Section \ref{sect:Whether}; those marks in the second column similarly refer to Section \ref{sect:Whom}. }
\label{tab:ResultsSummary}
\end{table}

Table \ref{tab:ResultsSummary}  very briefly summarizes our assessment.
Section \ref{sect:Whether} reports that the delegation rate is not lower when experts' payoffs are not shown, contrary to \autoref{h:CPP}(a) on the chasing past performance motive, 
while Section \ref{sect:Whom} reports that, consistent with \autoref{h:CPP}(b), delegation is indeed disproportionately to experts with highest payoffs when those payoffs are shown. 
The high delegation rates 
reported in Section \ref{sect:Whether} for Trivial tasks support our blame shifting \autoref{h:BlameShifting}.  
The even higher delegation rates for Complex tasks but not for Simple tasks contradict \autoref{h:DecisionCosts}(a) but support \autoref{h:DecisionCosts}(b).
The disproportionate choice of high quality experts reported in 
Section \ref{sect:Whom} supports \autoref{h:DecisionCosts}(c).
We found no evidence that would support \autoref{h:RiskTolerance} on risk tolerance. 
Regarding the supplementary \autoref{h:Price}, the evidence 
is mixed: delegation rates are price-sensitive only when experts' payoffs are not reported.

For us, the biggest surprise was how many of our investors  delegated---roughly half of them. 
The delegation frequencies are higher than those found elsewhere in the literature, e.g., 35\% in \citealt{apesteguia2020copy} and 17-38\% in \cite{holzmeister2022delegation}. We speculate that, in addition to other differences in the design and subject pool, our delegation frequencies arise in part from a fortuitous number of experts to choose from.
In contrast to our five experts, in \cite{apesteguia2020copy} subjects could choose among 80 leaders and in \cite{holzmeister2022delegation} there was no choice. 
Perhaps having no choice or too much choice may suppress delegation frequencies. 
Indeed, such an inverse U-shaped relationship is  documented for different choice domains; see e.g.\ \cite{iyengar2000choice}. 

Our experiment also uncovered some unexpected regularities. We saw very little sensitivity to the cost of delegation except when experts' own realized payment is not shown. Our interpretation is that showing realized payment distracts many of our investors, who give those payments more weight than they deserve and/or see even high cost as minor compared to payment variability.

Some null results also surprised us. 
Despite extensive specification search, we were unable to detect any connection between investors' characteristics (such as revealed risk aversion, choice quality and personality profile) and their decisions whether to delegate and if so, to whom. Evidently the propensity to delegate is less related to standard economic characteristics than one might have supposed.

Several caveats are in order. First, our results come from a specific subject pool. The on-line Prolific pool possibly resembles the subset of the investors who use social trading platforms, but it is not clear how closely it resembles the broader universe of investors, nor how much that matters. We are encouraged by the size and diversity of our Prolific subject pool, and by the fact that  delegation decisions seem unrelated to  subjects' demographic and psychological characteristics.

A further caveat is that our study was designed to investigate four specific motives for delegation.
Unsought findings --- e.g., that delegation propensity seems unrelated to revealed risk preferences or to standard personality traits ---  should be considered tentative until confirmed by new studies aimed at testing them. 

Moreover, in our experiment investors were only presented with a selected sample of experts, showcasing a conflict between low decision quality and high earnings on one hand and high decision quality and low earnings on the other. However, we believe that this selection is representative of  experts in a more general population. While in real world settings there may be a significant number of high quality experts, there surely is a much larger number of low quality potential experts and a few of them will have a lucky streak.
Thus, one should expect to see  low quality experts among the top earners. 

Delegation in our experiment entails replicating the decision  of the chosen expert, in some ways reminiscent of copy trading.\footnote{Copy trading in the field commits the follower to a leader's \textit{future} trades which may or may not resemble observed previous trades. In contrast, our delegation decision commits the follower to replicate a past decision of the leader, with a possibly different realized outcome.} 
In many field settings, delegation entails asking an expert to act on clients' behalf, tailoring choices to particular clients' circumstances rather than replicating (or even approximating) the expert's personal portfolio. Nonetheless experts often advise their clients to invest as they do personally  (\citealt{linnainmaa2021misguided}).\footnote{An interesting sidelight: \cite{bottasso2022higher} and \cite{holmen2023economic} show that finance professionals differ significantly in their risk preferences from the general population. Moreover, whilst they have superior decision quality compared to their clients when they are trading on their own, they are no better than their clients when choosing on their behalf \citep{stefan2022you}. } 
However, our focus is on how available information on experts systematically influences investors' choices of whether to delegate and if so to whom.
For our purposes, such tailored delegation would be counterproductive because we could no longer keep constant 
across investors the relevant attributes of experts. 



The current experiment gave investors little opportunity to learn from personal experience, since they saw their payoffs only at the very end. Future work might investigate whether the high fraction of investors who choose to delegate is stable over time as investors learn from experience about the pros and cons of delegation.

Another potential avenue for further research is to dissect blame shifting. 
For us, that motive is broadly construed to cover all sorts of non-instrumental decision avoidance.
Besides narrowly construed blame shifting, that motive might include potentially separable motives such as cognitive dissonance, regret aversion and perhaps others. 
Since broadly construed blame shifting turned out to be so important in our data, it might be worthwhile in future work to try to partition it into identifiable components that can be studied theoretically and empirically. 

From a policy point of view, the high fraction of investors who choose to delegate for seemingly welfare  reducing motives (especially blame shifting) appears alarming and raises questions about the optimal regulatory approach towards social trading platforms. 
On a more positive note, it is encouraging that a large fraction of investors chooses high quality experts when decision quality information is available.
That fact suggests that copy trading and other delegation platforms may wish to display more prominently information on experts' decision quality.\footnote{We acknowledge that it is a challenge to find appropriate quality measures in a field environment and to present them in a transparent manner.}


\clearpage
\bibliographystyle{plainnat}
\bibliography{refs}

\clearpage
\appendix
\renewcommand{\thetable}{\Alph{section}.\arabic{table}}
\setcounter{table}{0}

\section{Supplementary Data Analysis (Online)}
Here we offer additional empirical analysis, paralleling the material presented in Section \ref{sec:Results} and its subsections. 

\subsection{Whether to delegate?}

\begin{table}[ht]
\begin{center}
\resizebox{\linewidth}{!}{
\begin{tabular}{l | c c c c | c c c c}
\toprule
 & \multicolumn{4}{c|}{\bf logit model}
 & \multicolumn{4}{c}{\bf linear probability model} \\ \
 & (1) & (2) & (3) & (4) & (5) & (6) & (7) & (8) \\
\hline
Constant             & $-0.01$      & $0.02$       & $-0.09$      & $-0.05$      & $0.50^{***}$ & $0.51^{***}$ & $0.48^{***}$ & $0.49^{***}$ \\
                        & $(0.15)$     & $(0.15)$     & $(0.18)$     & $(0.18)$     & $(0.04)$     & $(0.04)$     & $(0.05)$     & $(0.05)$     \\
Trivial task                 &              & $-0.03$      &              & $-0.03$      &              & $-0.01$      &              & $-0.01$      \\
                        &              & $(0.11)$     &              & $(0.12)$     &              & $(0.03)$     &              & $(0.03)$     \\
Simple  task                & $0.03$       &              & $0.03$       &              & $0.01$       &              & $0.01$       &              \\
                        & $(0.11)$     &              & $(0.12)$     &              & $(0.03)$     &              & $(0.03)$     &              \\
Complex task                 & $0.46^{***}$ & $0.43^{***}$ & $0.46^{***}$ & $0.43^{***}$ & $0.11^{***}$ & $0.10^{***}$ & $0.11^{***}$ & $0.10^{***}$ \\
                        & $(0.12)$     & $(0.12)$     & $(0.12)$     & $(0.12)$     & $(0.03)$     & $(0.03)$     & $(0.03)$     & $(0.03)$     \\
First block                   & $0.04$       & $0.04$       & $0.04$       & $0.04$       & $0.01$       & $0.01$       & $0.01$       & $0.01$       \\
                        & $(0.11)$     & $(0.11)$     & $(0.11)$     & $(0.11)$     & $(0.03)$     & $(0.03)$     & $(0.03)$     & $(0.03)$     \\
Second block                  & $0.01$       & $0.01$       & $0.01$       & $0.01$       & $0.00$       & $0.00$       & $0.00$       & $0.00$       \\
                        & $(0.12)$     & $(0.12)$     & $(0.12)$     & $(0.12)$     & $(0.03)$     & $(0.03)$     & $(0.03)$     & $(0.03)$     \\
NoPay               & $0.12$       & $0.12$       & $0.39^{*}$   & $0.39^{*}$   & $0.03$       & $0.03$       & $0.10$       & $0.10$       \\
                        & $(0.12)$     & $(0.12)$     & $(0.20)$     & $(0.20)$     & $(0.03)$     & $(0.03)$     & $(0.05)$     & $(0.05)$     \\
NoQual               & $0.16$       & $0.16$       & $0.11$       & $0.11$       & $0.04$       & $0.04$       & $0.03$       & $0.03$       \\
                        & $(0.12)$     & $(0.12)$     & $(0.19)$     & $(0.19)$     & $(0.03)$     & $(0.03)$     & $(0.05)$     & $(0.05)$     \\
High Costs             & $-0.16$      & $-0.16$      & $-0.02$      & $-0.02$      & $-0.04$      & $-0.04$      & $-0.01$      & $-0.01$      \\
                        & $(0.10)$     & $(0.10)$     & $(0.19)$     & $(0.19)$     & $(0.02)$     & $(0.02)$     & $(0.05)$     & $(0.05)$     \\

NoPay $\times$ High Costs   &              &              & $-0.54^{*}$  & $-0.54^{*}$  &              &              & $-0.13^{*}$  & $-0.13^{*}$  \\
                        &              &              & $(0.26)$     & $(0.26)$     &              &              & $(0.06)$     & $(0.06)$     \\
NoQual $\times$ High Costs  &              &              & $0.11$       & $0.11$       &              &              & $0.03$       & $0.03$       \\
                        &              &              & $(0.25)$     & $(0.25)$     &              &              & $(0.06)$     & $(0.06)$     \\
\midrule
Num. obs.               & $1767$       & $1767$       & $1767$       & $1767$       & $1767$       & $1767$       & $1767$       & $1767$       \\
\bottomrule
\multicolumn{9}{l}{\scriptsize{$^{***}p<0.001$; $^{**}p<0.01$; $^{*}p<0.05$}}
\end{tabular}
}
\caption{
Regression results for delegation decision coefficients with standard errors in parenthesis.
Observation level: delegation decision (3 per subject). All standard errors are clustered at the individual level.
}
\label{tab:FullDelegationRegression}
\end{center}
\end{table}

\noindent
Table \ref{tab:FullDelegationRegression} presents an extended version of Table \ref{tab:DelegationGrandRegressionS}, including alternative specifications of the linear probability model as well as a logit model.
We again use both trivial and simple tasks as the baseline to ensure that the difference is significant between delegation in simple and complex tasks as well as between trivial and complex tasks.
The columns of the Table all confirm the robustness of results presented in Section \ref{sec:Results}.

\begin{table}[htbp]
\begin{center}
\resizebox{.85\linewidth}{!}{
\begin{tabular}{l c c c c c c}
\toprule
 & (1) & (2) & (3) & (4) & (5) & (6) \\
\midrule
Constant             & $1.57^{***}$ & $2.09^{***}$ & $2.08^{***}$ & $1.68^{***}$ & $2.24^{***}$ & $2.21^{***}$ \\
                        & $(0.09)$     & $(0.31)$     & $(0.31)$     & $(0.29)$     & $(0.41)$     & $(0.41)$     \\
NoPay               & $0.29^{*}$   & $0.31^{*}$   & $0.31^{*}$   & $0.29^{*}$   & $0.31^{*}$   & $0.31^{*}$   \\
                        & $(0.13)$     & $(0.13)$     & $(0.13)$     & $(0.13)$     & $(0.13)$     & $(0.13)$     \\
NoQual               & $0.08$       & $0.09$       & $0.10$       & $0.08$       & $0.09$       & $0.10$       \\
                        & $(0.13)$     & $(0.13)$     & $(0.13)$     & $(0.13)$     & $(0.13)$     & $(0.13)$     \\
High Costs             & $-0.02$      & $0.01$       & $0.02$       & $0.01$       & $0.04$       & $0.05$       \\
                        & $(0.13)$     & $(0.13)$     & $(0.13)$     & $(0.13)$     & $(0.13)$     & $(0.13)$     \\
NoPay $\times$ High Costs   & $-0.39^{*}$  & $-0.40^{*}$  & $-0.40^{*}$  & $-0.39^{*}$  & $-0.40^{*}$  & $-0.39^{*}$  \\
                        & $(0.18)$     & $(0.18)$     & $(0.18)$     & $(0.18)$     & $(0.18)$     & $(0.18)$     \\
NoQual $\times$ High Costs   & $0.08$       & $0.06$       & $0.04$       & $0.07$       & $0.05$       & $0.03$       \\
                        & $(0.18)$     & $(0.18)$     & $(0.18)$     & $(0.18)$     & $(0.18)$     & $(0.18)$     \\
Efficiency (CCEI)       &              & $-0.01$      & $-0.01$      &              & $-0.01$      & $-0.01$      \\
                        &              & $(0.00)$     & $(0.00)$     &              & $(0.00)$     & $(0.00)$     \\
Risk Tolerance            &              & $-0.07$      &              &              & $-0.06$      &              \\
                        &              & $(0.04)$     &              &              & $(0.04)$     &              \\
Relative Risk Aversion &              &              & $0.02$       &              &              & $0.02$       \\
                        &              &              & $(0.02)$     &              &              & $(0.02)$     \\
Big 5 (trusting)          &              &              &              & $-0.02$      & $-0.02$      & $-0.02$      \\
                        &              &              &              & $(0.04)$     & $(0.04)$     & $(0.04)$     \\
Big 5 (blameshifting)      &              &              &              & $-0.04$      & $-0.05$      & $-0.05$      \\
                        &              &              &              & $(0.04)$     & $(0.04)$     & $(0.04)$     \\
Big 5 (lazy)              &              &              &              & $0.04$       & $0.04$       & $0.04$       \\
                        &              &              &              & $(0.03)$     & $(0.03)$     & $(0.03)$     \\
Big 5 (outgoing)          &              &              &              & $-0.02$      & $-0.01$      & $-0.02$      \\
                        &              &              &              & $(0.04)$     & $(0.04)$     & $(0.04)$     \\
Big 5 (reserved)          &              &              &              & $0.01$       & $0.01$       & $0.01$       \\
                        &              &              &              & $(0.04)$     & $(0.04)$     & $(0.04)$     \\
\hline
Num. obs.               & $589$        & $589$        & $589$        & $589$        & $589$        & $589$        \\
\bottomrule
\multicolumn{7}{l}{\scriptsize{$^{***}p<0.001$; $^{**}p<0.01$; $^{*}p<0.05$}}
\end{tabular}
}
\caption{
OLS regression for delegation intensity coefficients with standard errors in parenthesis.
Observation level: delegated 0 to 3 times in the experiment (1 per subject).}
\label{tab:DelegationIntensity}
\end{center}
\end{table}

Table \ref{tab:DelegationIntensity} extends Table \ref{tab:DelegationIntensityS} from Section \ref{sec:Results}.
The table includes an additional measure of risk preferences,  \textit{risk tolerance}, which
is constructed exactly as in the expert risk ratings. 
That is, risk tolerance takes values of Low, Medium, and High based on where an investor's average implied coefficient of risk aversion falls in the distribution of the 60 potential experts. 
To control for the possible impact of personality traits,
the Table also includes responses to standard ``Big 5'' questions.
However, we detect no significant correlation between delegation intensity and efficiency, risk, or Big 5 traits.

\subsection{Whom to delegate?}
In this section, we focus on two (out of three) characteristics of the experts: quality and risk.
Recall that experts are chosen such that the earnings of Low quality experts are higher than those of High quality, and earnings increasing with increasing risk tolerance.
Along the dimension of quality, we separate experts into High (A, B, C) and Low (D,E) quality experts.
We are in particular, interested whether/when investors would like to delegate their decisions to the experts with Low quality ratings.
We code Low quality as 0 and High quality as 1.
For the analysis of the risk tolerance of experts, we code Low as 1 (expert A), Medium as 2 (experts B and D), and High as 3 (experts C and E).


\begin{table}[htbp]
\begin{center}
\resizebox{\linewidth}{!}{
\begin{tabular}{l| cccc | cccc}
\toprule
 & \multicolumn{4}{c|}{\bf OLS} & \multicolumn{4}{c}{\bf logit} \\
 & (1) & (2) & (3) & (4) & (5) & (6) & (7) & (8) \\
\midrule
Constant & $0.52^{***}$  & $0.57^{***}$  & $0.05$       & $0.03$       & $0.18$        & $0.34$        & $-3.01^{***}$ & $-3.18^{***}$ \\
                        & $(0.04)$      & $(0.06)$      & $(0.03)$     & $(0.03)$     & $(0.26)$      & $(0.28)$      & $(0.33)$      & $(0.39)$      \\
Trivial task            &               &               & $0.02$       & $0.02$       &               &               & $0.17$        & $0.18$        \\
                        &               &               & $(0.03)$     & $(0.03)$     &               &               & $(0.22)$      & $(0.22)$      \\
Simple task             & $-0.02$       & $-0.02$       &              &              & $-0.17$       & $-0.18$       &               &               \\
                        & $(0.03)$      & $(0.03)$      &              &              & $(0.22)$      & $(0.22)$      &               &               \\
Complex task            & $-0.03$       & $-0.03$       & $-0.00$      & $-0.00$      & $-0.21$       & $-0.21$       & $-0.03$       & $-0.02$       \\
                        & $(0.03)$      & $(0.03)$      & $(0.03)$     & $(0.03)$     & $(0.21)$      & $(0.21)$      & $(0.21)$      & $(0.21)$      \\
First block             & $0.02$        & $0.02$        & $0.02$       & $0.02$       & $0.17$        & $0.16$        & $0.17$        & $0.16$        \\
                        & $(0.03)$      & $(0.03)$      & $(0.03)$     & $(0.03)$     & $(0.22)$      & $(0.22)$      & $(0.22)$      & $(0.22)$      \\
Second block            & $-0.00$       & $-0.00$       & $-0.00$      & $-0.00$      & $-0.02$       & $-0.03$       & $-0.02$       & $-0.03$       \\
                        & $(0.03)$      & $(0.03)$      & $(0.03)$     & $(0.03)$     & $(0.22)$      & $(0.22)$      & $(0.22)$      & $(0.22)$      \\
AllInfo                    &               &               & $0.45^{***}$ & $0.52^{***}$ &               &               & $3.02^{***}$  & $3.34^{***}$  \\
                        &               &               & $(0.04)$     & $(0.05)$     &               &               & $(0.30)$      & $(0.41)$      \\
NoQual               & $0.43^{***}$  & $0.34^{***}$  & $0.87^{***}$ & $0.86^{***}$ & $2.46^{***}$  & $1.95^{***}$  & $5.48^{***}$  & $5.29^{***}$  \\
                        & $(0.04)$      & $(0.06)$      & $(0.02)$     & $(0.03)$     & $(0.25)$      & $(0.35)$      & $(0.34)$      & $(0.44)$      \\
NoPay               & $-0.45^{***}$ & $-0.52^{***}$ &              &              & $-3.02^{***}$ & $-3.34^{***}$ &               &               \\
                        & $(0.04)$      & $(0.05)$      &              &              & $(0.30)$      & $(0.41)$      &               &               \\
High Costs             & $-0.03$       & $-0.14^{*}$   & $-0.03$      & $0.01$       & $-0.26$       & $-0.57^{*}$   & $-0.26$       & $0.13$        \\
                        & $(0.02)$      & $(0.07)$      & $(0.02)$     & $(0.02)$     & $(0.21)$      & $(0.27)$      & $(0.21)$      & $(0.50)$      \\
NoQual $\times$ High Costs   &               & $0.18^{*}$    &              & $0.03$       &               & $1.04^{*}$    &               & $0.34$        \\
                        &               & $(0.07)$      &              & $(0.04)$     &               & $(0.50)$      &               & $(0.66)$      \\
NoPay $\times$ High Costs   &               & $0.15^{*}$    &              &              &               & $0.70$        &               &               \\
                        &               & $(0.07)$      &              &              &               & $(0.57)$      &               &               \\
AllInfo $\times$ High Costs      &               &               &              & $-0.15^{*}$  &               &               &               & $-0.70$       \\
                        &               &               &              & $(0.07)$     &               &               &               & $(0.57)$      \\
\midrule
Num. obs.               & $962$         & $962$         & $962$        & $962$        & $962$         & $962$         & $962$         & $962$         \\ \bottomrule
\multicolumn{9}{l}{\scriptsize{$^{***}p<0.001$; $^{**}p<0.01$; $^{*}p<0.05$}}
\end{tabular}
}
\caption{Choosing low-quality expert.
Results from OLS and logit regressions, coefficients with standard errors in parenthesis.}
\label{tab:LowQualityExperts}
\end{center}
\end{table}

Table \ref{tab:LowQualityExperts} presents the results of regression analysis of whether the investors choose to delegate their choice to the experts of Low quality rating.
Regression analysis confirms the findings evident from Figure \ref{fig:ExpertChosenByInfo}.
That is, we see that the investors take the decision based on the information available.
Recall that the low-quality experts are those with higher earnings.
In particular, if they do not have a quality rating for the expert, then they are significantly more likely to delegate their decision to the low-quality experts, 
and if they don't have the earnings information, then they are significantly less likely to delegate the decision to the low-quality experts.
We also see that investors treat both quality and payment information as relevant since even in the AllInfo condition they are still likely to delegate their decision to low-quality experts.
For the complexity and cost dimensions, we do not see robust effects.

\begin{table}[htbp]
\begin{center}
\resizebox{.85\linewidth}{!}{
\begin{tabular}{l|c c c c c c}
\hline
 & (1) & (2) & (3) & (4) & (5) & (6) \\
\hline
Consntant             & $2.45^{***}$  & $2.53^{***}$  & $2.44^{***}$  & $2.44^{***}$  & $2.47^{***}$  & $2.47^{***}$  \\
                        & $(0.08)$      & $(0.29)$      & $(0.08)$      & $(0.08)$      & $(0.08)$      & $(0.08)$      \\
Risk tolerance (task) & $0.02$        & $0.02$        &               &               &               &               \\
                        & $(0.04)$      & $(0.04)$      &               &               &               &               \\
Complex task                 & $0.00$        & $0.01$        & $0.03$        & $0.10$        & $0.03$        & $0.08$        \\
                        & $(0.05)$      & $(0.07)$      & $(0.05)$      & $(0.32)$      & $(0.05)$      & $(0.32)$      \\
First block                   & $-0.05$       & $-0.04$       & $-0.01$       & $-0.00$       & $-0.01$       & $-0.01$       \\
                        & $(0.07)$      & $(0.07)$      & $(0.06)$      & $(0.06)$      & $(0.06)$      & $(0.06)$      \\
Second block                  & $-0.01$       & $-0.01$       & $0.01$        & $0.01$        & $0.01$        & $0.01$        \\
                        & $(0.06)$      & $(0.06)$      & $(0.05)$      & $(0.05)$      & $(0.05)$      & $(0.05)$      \\
NoQual               & $-0.05$       & $-0.05$       & $-0.05$       & $-0.05$       & $-0.06$       & $-0.06$       \\
                        & $(0.07)$      & $(0.07)$      & $(0.07)$      & $(0.07)$      & $(0.07)$      & $(0.07)$      \\
NoPay               & $-0.34^{***}$ & $-0.34^{***}$ & $-0.45^{***}$ & $-0.45^{***}$ & $-0.45^{***}$ & $-0.45^{***}$ \\
                        & $(0.09)$      & $(0.09)$      & $(0.09)$      & $(0.09)$      & $(0.09)$      & $(0.09)$      \\
High costs             & $-0.01$       & $-0.01$       & $-0.09$       & $-0.09$       & $-0.09$       & $-0.09$       \\
                        & $(0.10)$      & $(0.10)$      & $(0.09)$      & $(0.09)$      & $(0.09)$      & $(0.09)$      \\
NoQual $\times$ high costs   & $0.03$        & $0.02$        & $0.05$        & $0.05$        & $0.06$        & $0.06$        \\
                        & $(0.11)$      & $(0.11)$      & $(0.10)$      & $(0.10)$      & $(0.10)$      & $(0.10)$      \\
NoPay $\times$ high cost   & $-0.02$       & $-0.02$       & $0.16$        & $0.16$        & $0.16$        & $0.16$        \\
                        & $(0.14)$      & $(0.14)$      & $(0.12)$      & $(0.12)$      & $(0.12)$      & $(0.12)$      \\
CCEI              &               & $-0.00$       &               & $-0.00$       &               & $-0.00$       \\
                        &               & $(0.00)$      &               & $(0.00)$      &               & $(0.00)$      \\
Risk tolerance (total)             &               &               & $0.03$        & $0.03$        &               &               \\
                        &               &               & $(0.03)$      & $(0.03)$      &               &               \\
Simple task                  &               &               & $0.03$        & $0.09$        & $0.02$        & $0.07$        \\
                        &               &               & $(0.05)$      & $(0.28)$      & $(0.05)$      & $(0.28)$      \\
Relative Risk Aversion         &               &               &               &               & $-0.01$       & $-0.01$       \\
                        &               &               &               &               & $(0.01)$      & $(0.01)$      \\
\hline
Num. obs.               & $665$         & $665$         & $962$         & $962$         & $962$         & $962$         \\
\hline
\multicolumn{7}{l}{\scriptsize{$^{***}p<0.001$; $^{**}p<0.01$; $^{*}p<0.05$}}
\end{tabular}
}
\caption{Risk tolerance of the expert. Results of OLS regression, coefficient with standard errors in parenthesis.}
\label{tab:RiskTolerance}
\end{center}
\end{table}

Table \ref{tab:RiskTolerance} presents the results of regression analysis for the risk tolerance of experts chosen in simple and complex tasks (tasks where risk communication might be a channel).
In addition to the risk preference measures used before we also include the task-level risk tolerance estimated since we restrict our attention to simple and complex tasks.
However, we find no correlation between the risk preferences of the investor and the risk tolerance of the expert they have chosen.
The investor's risk tolerance is also independent of the complexity of the task and the cost of delegation.

We do confirm a finding illustrated by Figure \ref{fig:ExpertChosenByInfo} that subjects prefer experts of at least Medium risk tolerance since the constant is about 2.5, that is subjects do not tend to choose the low-quality expert.
Another relevant finding is that is that in NoPay condition, subjects are choosing experts of lower risk tolerance.
Recall that, among the chosen experts, higher risk tolerance is associated with higher displayed earnings.
However, even in NoPay condition, the average risk tolerance of the expert is still about 2 (corresponding to Medium risk tolerance).
That is, even when we eliminate the chasing past performance motive, subjects still prefer to delegate to at least Medium risk tolerant experts.
The Table also confirms that subjects are sensitive to earnings information when choosing whom to delegate.

\begin{table}[ht]
\begin{center}
\resizebox{.85\linewidth}{!}{
\begin{tabular}{l c c c c c c}
\toprule
 & (1) & (2) & (3) & (4) & (5) & (6) \\
\hline
Constant             & $0.77^{***}$  & $0.91^{***}$  & $0.90^{***}$  & $0.87^{***}$  & $1.02^{***}$  & $1.02^{***}$  \\
                        & $(0.03)$      & $(0.11)$      & $(0.11)$      & $(0.10)$      & $(0.14)$      & $(0.14)$      \\
NoPay               & $0.18^{***}$  & $0.19^{***}$  & $0.19^{***}$  & $0.18^{***}$  & $0.19^{***}$  & $0.19^{***}$  \\
                        & $(0.04)$      & $(0.04)$      & $(0.04)$      & $(0.04)$      & $(0.04)$      & $(0.04)$      \\
NoQual               & $0.18^{***}$  & $0.19^{***}$  & $0.19^{***}$  & $0.18^{***}$  & $0.18^{***}$  & $0.19^{***}$  \\
                        & $(0.04)$      & $(0.04)$      & $(0.04)$      & $(0.04)$      & $(0.04)$      & $(0.04)$      \\
High costs             & $0.14^{**}$   & $0.15^{***}$  & $0.15^{***}$  & $0.15^{***}$  & $0.16^{***}$  & $0.16^{***}$  \\
                        & $(0.04)$      & $(0.04)$      & $(0.04)$      & $(0.04)$      & $(0.04)$      & $(0.04)$      \\
NoPay $\times$ High costs   & $-0.24^{***}$ & $-0.24^{***}$ & $-0.24^{***}$ & $-0.23^{***}$ & $-0.23^{***}$ & $-0.23^{***}$ \\
                        & $(0.06)$      & $(0.06)$      & $(0.06)$      & $(0.06)$      & $(0.06)$      & $(0.06)$      \\
NoQual $\times$ High costs   & $-0.18^{**}$  & $-0.19^{**}$  & $-0.19^{**}$  & $-0.18^{**}$  & $-0.19^{**}$  & $-0.19^{**}$  \\
                        & $(0.06)$      & $(0.06)$      & $(0.06)$      & $(0.06)$      & $(0.06)$      & $(0.06)$      \\
Efficiency              &               & $-0.00$       & $-0.00$       &               & $-0.00$       & $-0.00$       \\
                        &               & $(0.00)$      & $(0.00)$      &               & $(0.00)$      & $(0.00)$      \\
Risk tolerance             &               & $-0.01$       &               &               & $-0.01$       &               \\
                        &               & $(0.02)$      &               &               & $(0.02)$      &               \\
Relative Risk Aversion            &               &               & $0.00$        &               &               & $0.00$        \\
                        &               &               & $(0.01)$      &               &               & $(0.01)$      \\
Big 5 (trusting)          &               &               &               & $0.01$        & $0.01$        & $0.01$        \\
                        &               &               &               & $(0.01)$      & $(0.01)$      & $(0.01)$      \\
Big 5 (blameshifting)    &               &               &               & $-0.02$       & $-0.02$       & $-0.02$       \\
                        &               &               &               & $(0.01)$      & $(0.01)$      & $(0.01)$      \\
Big 5 (lazy)              &               &               &               & $0.01$        & $0.01$        & $0.01$        \\
                        &               &               &               & $(0.01)$      & $(0.01)$      & $(0.01)$      \\
Big 5 (outgoing)          &               &               &               & $-0.02$       & $-0.02$       & $-0.02$       \\
                        &               &               &               & $(0.01)$      & $(0.01)$      & $(0.01)$      \\
Big 5 (reserved)         &               &               &               & $-0.01$       & $-0.01$       & $-0.01$       \\
                        &               &               &               & $(0.01)$      & $(0.01)$      & $(0.01)$      \\
\hline
Num. obs.               & $589$         & $589$         & $589$         & $589$         & $589$         & $589$         \\
\bottomrule
\multicolumn{7}{l}{\scriptsize{$^{***}p<0.001$; $^{**}p<0.01$; $^{*}p<0.05$}}
\end{tabular}
}
\caption{OLS regression for delegated at least once, coefficient with standard errors in parenthesis. Observation level: delegated at least one in the experiment (1 per subject).}
\label{tab:DelegatedOnceOLS}
\end{center}
\end{table}

\subsection{Who delegates?}
To investigate the difference between investors who decided to delegate and those who didn't, we construct a variable that is equal to 1 if the investor decided to delegate at least once (among three tasks).
Table \ref{tab:DelegatedOnceOLS} presents regression results for this
dependent variable.
All results presented are robust to changing the specification to logit or probit.
There is no significant correlation with efficiency, risk, or personality traits.
Thus in conjunction with the results from Table \ref{tab:DelegationIntensity} none of the individual characteristics can explain whether the investor wants to delegate the decision or how frequently they would like to delegate the decision.

However, we do observe the effects of informational conditions on the delegation.
In particular, we see that the presence of conflicting information reduces the delegation.
If we remove either the information about the quality or about earnings of the experts, then the investor 18 percentage points is more likely to delegate
in the case of low costs of delegation.
For the high costs treatment, even in the presence of conflicting information investors are more likely to delegate at least once.
Note that these results cannot be translated to the intensity of delegation.

\subsection{Consistency of Decisions}

\begin{table}[ht]
\centering
\resizebox{1\linewidth}{!}{
\begin{tabular}{l|cccc}
  \toprule
 & Efficiency & Low Quality Expert & Strong Consistency & Weak Consistency \\ 
  \midrule
Efficiency   & 1.00 & 0.02 & 0.16 & 0.21  \\ 
Low Quality Expert & 0.02 & 1.00 & 0.00 & 0.09  \\ 
Strong Consistency  & 0.16 & 0.00 & 1.00 & 0.53 \\ 
Weak Consistency & 0.21 & 0.09 & 0.53 & 1.00 \\ 
   \bottomrule
\end{tabular}
}
\caption{Correlation between different quality indicators. Data from simple and complex tasks of all information treatment.}
\label{tab:CorrelationsConsistency}
\end{table}

\noindent
Table \ref{tab:CorrelationsConsistency} presents the results correlating different measures of decision-making quality.
Efficiency is the Afriat efficiency in the investment task.
Low Quality Expert is a dummy variable for the subject choosing a low quality expert.
Weak Consistency is a dummy variable that a subject who delegated in trivial or simple task also delegated in the complex task.
Strong Consistency indicates that if subject delegated in trivial task, then they delegated in simple and complex task, and if they delegated in simple task, then they delegated in complex task.
We do see that none of the efficiency or consistency measures is correlated with the choice of the low quality expert.
However, we do see the correlation between the investment-related and delegated-related efficiency measures.

\subsection{Robustness: Consistent Subjects}
In this section we present the regression analysis related to the analysis of consistent subjects.
That is subjects who delegated in simple and complex tasks if they delegated in trivial task and who delegated in complex task if they delegated in the simple task.

\begin{table}[htb]
\begin{center}
\resizebox{1\linewidth}{!}{
\begin{tabular}{l c c c c | c c c c}
\hline
& 
\multicolumn{4}{c|}{\bf logit model}
& 
\multicolumn{4}{c}{\bf linear probability model} \\
 & (1) & (2) & (3) & (4) & (5) & (6) & (7) & (8) \\
\hline
Constant    & $-0.78^{***}$ & $0.31$        & $-0.90^{**}$ & $0.21$        & $0.32^{***}$ & $0.57^{***}$  & $0.30^{***}$ & $0.56^{***}$ \\
                        & $(0.22)$      & $(0.22)$      & $(0.27)$     & $(0.26)$      & $(0.05)$     & $(0.05)$      & $(0.06)$     & $(0.05)$     \\
Trivial task      &               & $-1.09^{***}$ &              & $-1.11^{***}$ &              & $-0.26^{***}$ &              &              \\
                        &               & $(0.11)$      &              & $(0.11)$      &              & $(0.02)$      &              &              \\
Simple task     & $1.09^{***}$  &               & $1.11^{***}$ &               & $0.26^{***}$ &               & $0.26^{***}$ &              \\
                        & $(0.11)$      &               & $(0.11)$     &               & $(0.02)$     &               & $(0.02)$     &              \\
Complex task      & $2.31^{***}$  & $1.22^{***}$  & $2.34^{***}$ & $1.24^{***}$  & $0.51^{***}$ & $0.25^{***}$  & $0.51^{***}$ &              \\
                        & $(0.15)$      & $(0.12)$      & $(0.15)$     & $(0.12)$      & $(0.03)$     & $(0.02)$      & $(0.03)$     &              \\
First block  & $0.08$        & $0.08$        & $0.10$       & $0.10$        & $0.02$       & $0.02$        & $0.02$       & $-0.00$      \\
                        & $(0.13)$      & $(0.13)$      & $(0.13)$     & $(0.13)$      & $(0.03)$     & $(0.03)$      & $(0.03)$     & $(0.03)$     \\
Second block  & $-0.09$       & $-0.09$       & $-0.09$      & $-0.09$       & $-0.02$      & $-0.02$       & $-0.02$      & $-0.03$      \\
                        & $(0.12)$      & $(0.12)$      & $(0.13)$     & $(0.13)$      & $(0.03)$     & $(0.03)$      & $(0.03)$     & $(0.03)$     \\
NoPay               & $0.19$        & $0.19$        & $0.62$       & $0.62$        & $0.04$       & $0.04$        & $0.12$       & $0.12$       \\
                        & $(0.24)$      & $(0.24)$      & $(0.35)$     & $(0.35)$      & $(0.05)$     & $(0.05)$      & $(0.07)$     & $(0.07)$     \\
NoQual     & $0.28$        & $0.28$        & $0.13$       & $0.13$        & $0.06$       & $0.06$        & $0.03$       & $0.03$       \\
                        & $(0.24)$      & $(0.24)$      & $(0.34)$     & $(0.34)$      & $(0.05)$     & $(0.05)$      & $(0.07)$     & $(0.07)$     \\
High Costs         & $-0.46^{*}$   & $-0.46^{*}$   & $-0.22$      & $-0.22$       & $-0.09^{*}$  & $-0.09^{*}$   & $-0.04$      & $-0.04$      \\
                        & $(0.19)$      & $(0.19)$      & $(0.35)$     & $(0.35)$      & $(0.04)$     & $(0.04)$      & $(0.07)$     & $(0.07)$     \\
NoPay $\times$ High Costs   &               &               & $-0.90$      & $-0.90$       &              &               & $-0.18$      & $-0.18$      \\
                        &               &               & $(0.48)$     & $(0.48)$      &              &               & $(0.10)$     & $(0.10)$     \\
NoQual $\times$ High Costs  &               &               & $0.21$       & $0.21$        &              &               & $0.04$       & $0.04$       \\
                        &               &               & $(0.47)$     & $(0.47)$      &              &               & $(0.10)$     & $(0.10)$     \\
\hline
Num. obs.               & $1020$        & $1020$        & $1020$       & $1020$        & $1020$       & $1020$        & $1020$       & $1020$       \\
\hline
\multicolumn{9}{l}{\scriptsize{$^{***}p<0.001$; $^{**}p<0.01$; $^{*}p<0.05$}}
\end{tabular}
}
\caption{Regression results for delegation decision for restricted sample of ``consistent'' subjects. Standard errors in parenthesis.
Observation level: delegation decision (3 per subject). All standard errors are clustered at the individual level.
}
\label{tab:FullRegressionConsistentSubjects}
\end{center}
\end{table}

Table \ref{tab:FullRegressionConsistentSubjects} presents the regression analysis for the consistent subjects.
Note that the increase in delegation with complexity is highly significant in this case due to the construction of the consistent subjects.
The main difference (not induced by construction) we get is the significance of the High Costs.
That is, consistent subjects are price sensitive while we do not observe this in the general sample.
However, this price sensitivity is not robust to controlling for the product of High costs with the information conditions.

\begin{table}[htb]
\begin{center}
\resizebox{\columnwidth}{!}{
\begin{tabular}{l c c c c c c}
\hline
 & (1) & (2) & (3) & (4) & (5) & (6) \\
\hline
Constant   & $0.70^{***}$ & $0.69^{***}$ & $0.67^{***}$ & $0.85^{**}$  & $0.79^{*}$   & $0.67^{*}$   \\
                        & $(0.05)$     & $(0.06)$     & $(0.05)$     & $(0.33)$     & $(0.33)$     & $(0.32)$     \\
First block    & $-0.07$      & $-0.01$      & $0.08$       & $-0.40$      & $-0.05$      & $0.62$       \\
                        & $(0.05)$     & $(0.05)$     & $(0.05)$     & $(0.33)$     & $(0.36)$     & $(0.38)$     \\
Second block   & $0.04$       & $-0.01$      & $-0.04$      & $0.36$       & $-0.04$      & $-0.23$      \\
                        & $(0.05)$     & $(0.05)$     & $(0.05)$     & $(0.39)$     & $(0.35)$     & $(0.33)$     \\
NoPay              & $0.22^{**}$  & $0.23^{***}$ & $0.22^{**}$  & $1.52^{**}$  & $1.62^{**}$  & $1.58^{**}$  \\
                        & $(0.07)$     & $(0.07)$     & $(0.07)$     & $(0.54)$     & $(0.53)$     & $(0.53)$     \\
NoQual    & $0.21^{**}$  & $0.22^{**}$  & $0.22^{**}$  & $1.36^{*}$   & $1.42^{**}$  & $1.45^{**}$  \\
                        & $(0.07)$     & $(0.07)$     & $(0.07)$     & $(0.54)$     & $(0.54)$     & $(0.54)$     \\
High Costs     & $0.13$       & $0.13$       & $0.14$       & $0.71$       & $0.71$       & $0.78$       \\
                        & $(0.07)$     & $(0.07)$     & $(0.07)$     & $(0.45)$     & $(0.45)$     & $(0.45)$     \\
NoPay $\times$ High Costs  & $-0.29^{**}$ & $-0.29^{**}$ & $-0.29^{**}$ & $-1.94^{**}$ & $-1.96^{**}$ & $-1.97^{**}$ \\
                        & $(0.10)$     & $(0.10)$     & $(0.10)$     & $(0.72)$     & $(0.72)$     & $(0.73)$     \\
Noqual $\times$ High Costs  & $-0.19$      & $-0.19$      & $-0.21^{*}$  & $-1.23$      & $-1.26$      & $-1.41$      \\
                        & $(0.10)$     & $(0.11)$     & $(0.10)$     & $(0.75)$     & $(0.75)$     & $(0.75)$     \\
\hline
Num. obs.               & $340$        & $340$        & $340$        & $340$        & $340$     & $340$        \\
\hline
\multicolumn{7}{l}{\scriptsize{$^{***}p<0.001$; $^{**}p<0.01$; $^{*}p<0.05$}}
\end{tabular}
}
\caption{
Whether subject delegated in at least one of the tasks.
}
\label{tab:DelegatedConsistentSubjects}
\end{center}
\end{table}

Table \ref{tab:DelegatedConsistentSubjects} presents the regression analysis for whether consistent subjects delegated at least once.
Models (1)-(3) are linear probability, models (4)-(6) are logits.
We do see that the results are consistent with those observed for the general sample.

\begin{table}[htb]
\begin{center}
\resizebox{\columnwidth}{!}{
\begin{tabular}{l|c c c c| c c c c}
\hline
& \multicolumn{4}{c}{\textbf{OLS}} & \multicolumn{4}{c}{\textbf{logit}} \\
 & (1) & (2) & (3) & (4) & (5) & (6) & (7) & (8) \\
\hline
Constant             & $0.56^{***}$  & $0.61^{***}$  & $0.05$       & $0.03$       & $0.18$        & $0.34$        & $-3.01^{***}$ & $-3.18^{***}$ \\
                        & $(0.06)$      & $(0.07)$      & $(0.03)$     & $(0.03)$     & $(0.26)$      & $(0.28)$      & $(0.32)$      & $(0.39)$      \\
Simple Task                  & $-0.05$       & $-0.04$       &              &              & $-0.17$       & $-0.18$       &               &               \\
                        & $(0.04)$      & $(0.04)$      &              &              & $(0.22)$      & $(0.22)$      &               &               \\
Complex Task                 & $-0.08^{*}$   & $-0.07$       & $-0.03$      & $-0.03$      & $-0.21$       & $-0.21$       & $-0.03$       & $-0.02$       \\
                        & $(0.04)$      & $(0.04)$      & $(0.03)$     & $(0.03)$     & $(0.21)$      & $(0.21)$      & $(0.22)$      & $(0.22)$      \\
Trivial Task                 &               &               & $0.05$       & $0.04$       &               &               & $0.17$        & $0.18$        \\
                        &               &               & $(0.04)$     & $(0.04)$     &               &               & $(0.22)$      & $(0.22)$      \\
First block                   & $0.03$        & $0.03$        & $0.03$       & $0.03$       & $0.17$        & $0.16$        & $0.17$        & $0.16$        \\
                        & $(0.03)$      & $(0.03)$      & $(0.03)$     & $(0.03)$     & $(0.21)$      & $(0.21)$      & $(0.21)$      & $(0.21)$      \\
Second Block                  & $0.02$        & $0.02$        & $0.02$       & $0.02$       & $-0.02$       & $-0.03$       & $-0.02$       & $-0.03$       \\
                        & $(0.03)$      & $(0.03)$      & $(0.03)$     & $(0.03)$     & $(0.22)$      & $(0.22)$      & $(0.22)$      & $(0.22)$      \\
AllInfo                    &               &               & $0.46^{***}$ & $0.54^{***}$ &               &               & $3.02^{***}$  & $3.34^{***}$  \\
                        &               &               & $(0.04)$     & $(0.06)$     &               &               & $(0.29)$      & $(0.41)$      \\
NoQual               & $0.43^{***}$  & $0.31^{***}$  & $0.89^{***}$ & $0.86^{***}$ & $2.46^{***}$  & $1.95^{***}$  & $5.48^{***}$  & $5.29^{***}$  \\
                        & $(0.05)$      & $(0.07)$      & $(0.03)$     & $(0.04)$     & $(0.25)$      & $(0.35)$      & $(0.34)$      & $(0.44)$      \\
NoPay               & $-0.46^{***}$ & $-0.54^{***}$ &              &              & $-3.02^{***}$ & $-3.34^{***}$ &               &               \\
                        & $(0.04)$      & $(0.06)$      &              &              & $(0.29)$      & $(0.41)$      &               &               \\
High Costs            & $-0.04$       & $-0.20^{*}$   & $-0.04$      & $0.02$       & $-0.26$       & $-0.57^{*}$   & $-0.26$       & $0.13$        \\
                        & $(0.03)$      & $(0.08)$      & $(0.03)$     & $(0.03)$     & $(0.21)$      & $(0.27)$      & $(0.21)$      & $(0.51)$      \\
NoQual $\times$ High Costs   &               & $0.26^{**}$   &              & $0.04$       &               & $1.04^{*}$    &               & $0.34$        \\
                        &               & $(0.09)$      &              & $(0.05)$     &               & $(0.50)$      &               & $(0.67)$      \\
NoPay $\times$ High Costs   &               & $0.21^{*}$    &              &              &               & $0.70$        &               &               \\
                        &               & $(0.08)$      &              &              &               & $(0.58)$      &               &               \\
AllInfo $\times$ High Costs     &               &               &              & $-0.21^{*}$  &               &               &               & $-0.70$       \\
                        &               &               &              & $(0.08)$     &               &               &               & $(0.58)$      \\
\hline
Num. obs.               & $568$         & $568$         & $568$        & $568$        & $962$         & $962$         & $962$         & $962$         \\
\hline
\multicolumn{9}{l}{\scriptsize{$^{***}p<0.001$; $^{**}p<0.01$; $^{*}p<0.05$}}
\end{tabular}
}
\caption{Choosing low-quality expert. Results from OLS and logit regressions,
coefficients with standard errors in parenthesis.}
\label{tab:ConsistentLowQualityExperts}
\end{center}
\end{table}

Table \ref{tab:ConsistentLowQualityExperts} presents the results for choosing low quality experts focusing on the subsample of consistent subjects.
The results are consistent with those of the general sample.
The only difference is that we observe that in one of the specifications consistent subjects are less likely to choose low-quality expert in complex task.
However, this results is not robust to alternative regression specifications.

\clearpage
\section{Details on CCEI (Online)}\label{App:CCEI}
We now explain how we adapt the
\cite{afriat1973system}
Critical Cost Efficiency Index (CCEI) to our setting.
Recall that we have a finite set of observations for each subject.
A subject is choosing from a finite budget set $B_t$, which is defined by the constant price vector $p_t$, such that
$$
x \in B_t \text{ if and only if } p_t x \le 1.
$$
(Figure \ref{fig:BudgetsForExperiment} illustrates all budget sets used in the experiment.)
Denote by $x_t$ the allocation chosen from the budget $B_t$.
Then, under the \textbf{Generalized Axiom of Revealed Preferences (GARP)} for every sequence $t=1,\ldots, n$ such that $p_{t+1} x_t \le 1$, it holds that $p_n x_1 \ge 1$.
It is well known \citep[see][]{varian1982nonparametric} that GARP is equivalent to the existence of a locally non-satiated continuous and concave utility function that generated the data.
Since we consider preferences consistent with stochastic dominance and we consider assets with only two equally likely states of the world, we can use a simplified version of FSD-GARP \citep[see][]{heufer2014nonparametric,castillo2018revealed}.
Let $\sigma((x_1,x_2)) = (x_2,x_1)$ be the permutation of the choice $x$.
A data set satisfies \textbf{FSD-GARP} if for every sequence $t=1,\ldots, n$ such that $p_{t+1} x_t \le 1$ or $p_{t+1} \sigma(x_t) \le 1$ it holds that $p_n x_1 \ge 1$ and $p_n \sigma(x_1) \ge 1$.
 FSD-GARP is then equivalent to the existence of a utility function that satisfies stochastic dominance and generated the data.
Finally, given the axiom we can define \textbf{CCEI} as the maximum $e\in [0,1]$ such that for every sequence $t=1,\ldots, n$ with $p_{t+1} x_t \le e$ or $p_{t+1} \sigma(x_t) \le e$, it holds that $p_n x_1 \ge e$ and $p_n \sigma(x_1) \ge e$.

Intuitively, a CCEI of $e$ means that the chosen bundle is only better than everything in the reduced budget set $p_t x \le e$ instead of the full budget set $p_t x\le 1$,
i.e., she is maximizing her preferences as if her income were only $e<1$.
In that sense, she is leaving a $1-e$ share of income on the table,
and could have attained a better bundle if she rationally deployed the 
$1-e$ share.

\section{Experimental Details (Online)}
\label{appendix:Instructions}

\subsection{Budget Specifications}

\input{paper_Figs_and_Tabs/budgets_alt}

\noindent
Figures \ref{fig:BudgetsForExperiment} presents the entire set of budgets for Trivial, Simple and Complex tasks.
Recall that Trivial and Simple tasks are menu choices, while the Complex task is the portfolio choice.
Thus, for the Complex task we also present the comprehensive closure of the budget set.
Every budget for Simple and Complex task contains two points that are dominated if subject is consistent with maximizing a utility function that satisfies first-order stochastic dominance.
These points are marked with red.
One dominated point for every budget set is strictly inside the budget set, so it is dominanted as long as subject has monotone utility.
Second dominated point for every budget set is on the "short side", that is, this choice is not dominated if subject has only monotone utility, but is dominated if subject has utility that satisfies first-order stochastic dominance. 

\subsection{Experimental Instructions}
Below we present the full set of experimental instructions.
General instructions been shown to every subject prior to the beginning of the experiment.
Investment tasks were shown in random order of \say{Blocks}.

\clearpage


\section*{Supplementary material}

# General Instrucitons:

## Overview

In this experiment you will face three blocks of five tasks each. Each task involves an economic decision. At the end of the experiment, one of the blocks will be randomly (with equal likelihood) selected for payment. Out of this block you will be paid for one out of your first four decisions (with equal likelihood), and you also will be paid for your last decision in that block.

You will certainly receive **£1.5 for participation**. You will also earn an additional £0-11 depending on the decisions you make and on the random events, as described below. Throughout the experiment we use tokens as currency. At the end of the experiment tokens will be converted at the rate of **20 tokens per £1**. In the beginning of the experiment you are given the **endowment of 20 tokens (£1)** and further earnings (and losses) will be added to the initial endowment.

Before making decisions in each block, you will receive instructions for that block and a couple of questions to check your understanding. Please read the instructions carefully because fully understanding them will help you make better decisions and take home more money.

# Trivial Investment Task:

# Instructions (Block 2 out of 3)

In this block, you will make five decisions. In each of these decisions, you will choose one of two investment options.

The table on the decision screen shows how the payoff of each option depends on the state of the world. There are two equally likely states of the world: Heads and Tails. The actual state will be determined by a virtual coin flip (using the computer's random number generator).

In the example decision below, if you choose Option 1, then you will receive 100 tokens if the coin comes up Heads, and will receive 0 tokens if the coin comes up Tails. If you choose Option 2, then you will receive 0 tokens if the coin comes up Heads, and 100 tokens if Tails.

|  | **Payment if Heads** | **Payment if Tails** |
|---|---|---|
| ○ **Option 1** | 100 | 0 |
| ○ **Option 2** | 0 | 100 |

# Simple Investment Task:

# Instructions (Block 3 out of 3)

In this block, you will make five decisions. In each of these decisions, you will choose one of six investment options.

The table on the decision screen shows how the payoff of each option depends on the state of the world. There are two equally likely states of the world: Heads and Tails. The actual state will be determined by a virtual coin flip (using the computer's random number generator).

In the example decision below, if you choose Option 1, then you will receive 55 tokens if the coin comes up Heads, and will receive 15 tokens if the coin comes up Tails. If instead you choose Option 3, then you will receive 40 tokens if the coin comes up Heads, and 30 tokens if Tails. Likewise, if you choose Option 4, then you will receive 88 tokens if the coin comes up Heads, and 6 tokens if Tails.

|  | Payment if Heads | Payment if Tails |
|---|---|---|
| ○ Option 1 | 55 | 15 |
| ○ Option 2 | 68 | 16 |
| ○ Option 3 | 40 | 30 |
| ○ Option 4 | 88 | 6 |
| ○ Option 5 | 20 | 40 |
| ○ Option 6 | 34 | 33 |

# Complex Investment Task:

# Instructions I (Block 1 out of 3)

In this block, you will make five decisions. In each of these decisions, you will distribute 100 units among six investment options. At the end of the experiment, one of the decisions will be randomly chosen and it will determine your payment.

The table on the decision screen shows how the payoff of each option depends on the state of the world. There are two equally likely states of the world: Heads and Tails. The actual state will be determined by a virtual coin flip (using the computer's random number generator).

|  | Payment if Heads | Payment if Tails |  |  |
|---|---|---|---|---|
| Option 1 | 100 | 0 |  | 0 |
| Option 2 | 50 | 10 |  | 50 |
| Option 3 | 20 | 16 |  | 0 |
| Option 4 | 30 | 14 |  | 50 |
| Option 5 | 60 | 4 |  | 0 |
| Option 6 | 10 | 18 |  | 0 |
| Leftover Units |  |  |  | 0 |
|  |  |  | Reset | Submit |

# Instructions II (Block 1 out of 3)

Your payoff each round will depend on how you choose to allocate 100 units among the available options, and also on the outcome of a virtual coin flip. In the example below, if you choose to allocate all 100 units in Option 1, then you would receive 100*100 = 10,000 mini-tokens = 100 tokens if the coin comes up Heads, and would receive 0 tokens if the coin comes up Tails. If you choose instead to invest 50 units in Option 2 and 50 units in Option 4, then you would receive 50*50+50*30 = 4000 mini-tokens = 40 tokens if the coin comes up Heads, and you would receive 50*10+50*14 = 1200 mini-tokens = 12 tokens if the coin comes up Tails.

|  | Payment if Heads | Payment if Tails |  |  |
|---|---|---|---|---|
| Option 1 | 100 | 0 |  | 0 |
| Option 2 | 50 | 10 |  | 50 |
| Option 3 | 20 | 16 |  | 0 |
| Option 4 | 30 | 14 |  | 50 |
| Option 5 | 60 | 4 |  | 0 |
| Option 6 | 10 | 18 |  | 0 |
| Leftover Units |  |  |  | 0 |
|  |  |  | Reset | Submit |

# Instructions III (Block 1 out of 3)

To distribute your 100 units, use your mouse to move the blue bars in the fourth column, or use the -/+ buttons in the last column. If you want to start over, just click the **"Reset"** button.

To finalize your distribution, first check that you have distributed all 100 units. (To avoid unnecessary waste, the program will not let you proceed if there are leftover units.) Then click the **"Submit"** button and, provided there are no leftover units, a **"Next"** button will appear. Clicking it will finalize your distribution on that task.

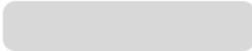

# Delegation instructions: All Info Treatment

## Instructions: Delegation

You can now choose to either delegate the task to a previous investor or else you can continue making your own choices as in the previous four rounds. If you delegate to a previous investor, then the computer will automatically choose for you exactly the same asset as that investor chose in that round.

As usual, your screen will display information you will need if you decide to make your own choices. It will also display a list of five investors who have previously played this game. The display also tells you something about each of those previous investors: the final payments they received, their risk ratings and their quality ratings.

**Their final payments**, like yours, come from the last round and one of the other four rounds in a single randomly chosen block. Please note that it is often from a different block (and round) than that for which you delegate.

**Their risk rating** --- High, Medium, or Low --- is based on all their choices during their session. High risk rating means they often made choices that result in a very high payment in one state and a very low payment in the other state, while low risk rating indicates mostly choices that give a moderate payout no matter what the state turns out to be.

**Their quality rating** --- High or Low --- indicates the "efficiency" of the investor's choices. A low rating indicates that they often could have obtained higher payments in both states if they had made different choices.

**Please take the delegation decision:**

- I do not want to delegate my decision

To delegate, please select one of the investors below (costs 20 tokens = £1.00)

| Investor | Quality Rating | Risk Rating | Earnings |
|---|---|---|---|
| A | High | Low | £2.5 |
| B | High | Medium | £2.9 |
| C | High | High | £3.5 |
| D | Low | Medium | £4.6 |
| E | Low | High | £8.5 |

Next

# Delegation instructions: No Payment Info Treatment

# Instructions: Delegation

You can now choose to either delegate the task to a previous investor or else you can continue making your own choices as in the previous four rounds. If you delegate to a previous investor, then the computer will automatically choose for you exactly the same asset as that investor chose in that round.

As usual, your screen will display information you will need if you decide to make your own choices. It will also display a list of five investors who have previously played this game. The display also tells you something about each of those previous investors: their risk ratings and their quality ratings.

**Their risk rating** --- High, Medium, or Low --- is based on all their choices during their session. High risk rating means they often made choices that result in a very high payment in one state and a very low payment in the other state, while low risk rating indicates mostly choices that give a moderate payout no matter what the state turns out to be.

**Their quality rating** --- High or Low --- indicates the "efficiency" of the investor's choices. A low rating indicates that they often could have obtained higher payments in both states if they had made different choices.

**Please take the delegation decision:**

○ I do not want to delegate my decision

To delegate, please select one of the investors below (costs 2 tokens = £0.1)

| | Investor | Quality Rating | Risk Rating |
|---|---|---|---|
| ○ | A | Low | Medium |
| ○ | B | High | Low |
| ○ | C | High | Medium |
| ○ | D | High | High |
| ○ | E | Low | High |

Next

# Delegation instructions: No Quality Info Treatment

You can now choose to either delegate the task to a previous investor or else you can continue making your own choices as in the previous four rounds. If you delegate to a previous investor, then the computer will automatically choose for you exactly the same asset as that investor chose in that round.

As usual, your screen will display information you will need if you decide to make your own choices. It will also display a list of five investors who have previously played this game. The display also tells you something about each of those previous investors: the final payments they received and their risk ratings.

**Their final payments**, like yours, come from the last round and one of the other four rounds in a single randomly chosen block. Please note that it is often from a different block (and round) than that for which you delegate.

**Their risk rating** --- High, Medium, or Low --- is based on all their choices during their session. High risk rating means they often made choices that result in a very high payment in one state and a very low payment in the other state, while low risk rating indicates mostly choices that give a moderate payout no matter what the state turns out to be.

**Please take the delegation decision:**

○ I do not want to delegate my decision

To delegate, please select one of the investors below (costs 2 tokens = £0.1)

| | Investor | Risk Rating | Earnings |
|---|---|---|---|
| ○ | A | Medium | £2.9 |
| ○ | B | High | £3.5 |
| ○ | C | High | £8.5 |
| ○ | D | Low | £2.5 |
| ○ | E | Medium | £4.6 |



\end{document}

%% file: paper_Figs_and_Tabs/treatment_effects.tex
\begin{figure}[h]
\centering
\begin{subfigure}[t]{0.49\linewidth}
\centering
\resizebox{1\linewidth}{!}{
\begin{tikzpicture}
\begin{axis}[
    xtick={1,3,5},
    xticklabels={Trivial, Simple, Complex},
    xmin = 0, xmax = 6,
    ymin = 0, ymax = 1,
    ylabel = {Frequency}
]

\addplot[ybar,
    fill=MainColor,
    MainColor,
    bar width = 1
     ]
    coordinates {
(1, .504)
(3, .513)
(5, .616)
};


\end{axis}
\end{tikzpicture}
}
\caption{Delegation by complexity}
\label{fig:complexRes}
\end{subfigure}
\hfill
\begin{subfigure}[t]{0.49\linewidth}
\centering
\resizebox{1\linewidth}{!}{
\begin{tikzpicture}
\begin{axis}[
    ybar,
    legend pos = north east,
    legend style={draw=none},
    legend cell align = left,
    xtick={1,3,5},
    xticklabels={\footnotesize AllInfo, \footnotesize  NoQual,\footnotesize NoPay},
    xmin = 0, xmax = 6,
    ymin = 0, ymax = 1,
    ylabel = {Frequency}
]


\addplot[ybar,
    fill=MainColor,
    MainColor,
    bar width = .4
     ]
    coordinates {
    (1, 0.5238095)
    (3, 0.5498282)
    (5, 0.6190476)
};
\addlegendentry{Low Cost};

\addplot[ybar,
    fill=SecondaryColor,
    SecondaryColor,
    bar width = .4
     ]
    coordinates {
    (1, 0.5185185)
    (3, 0.5723906)
    (5, 0.4829932)
};
\addlegendentry{High Cost};

\end{axis}
\end{tikzpicture}
}
\caption{Delegation by information and costs}
\label{fig:costbycomplexRes}    
\end{subfigure}
\caption{
Delegation frequencies.
}
\end{figure}

%% file: paper_Figs_and_Tabs/budgets_alt.tex
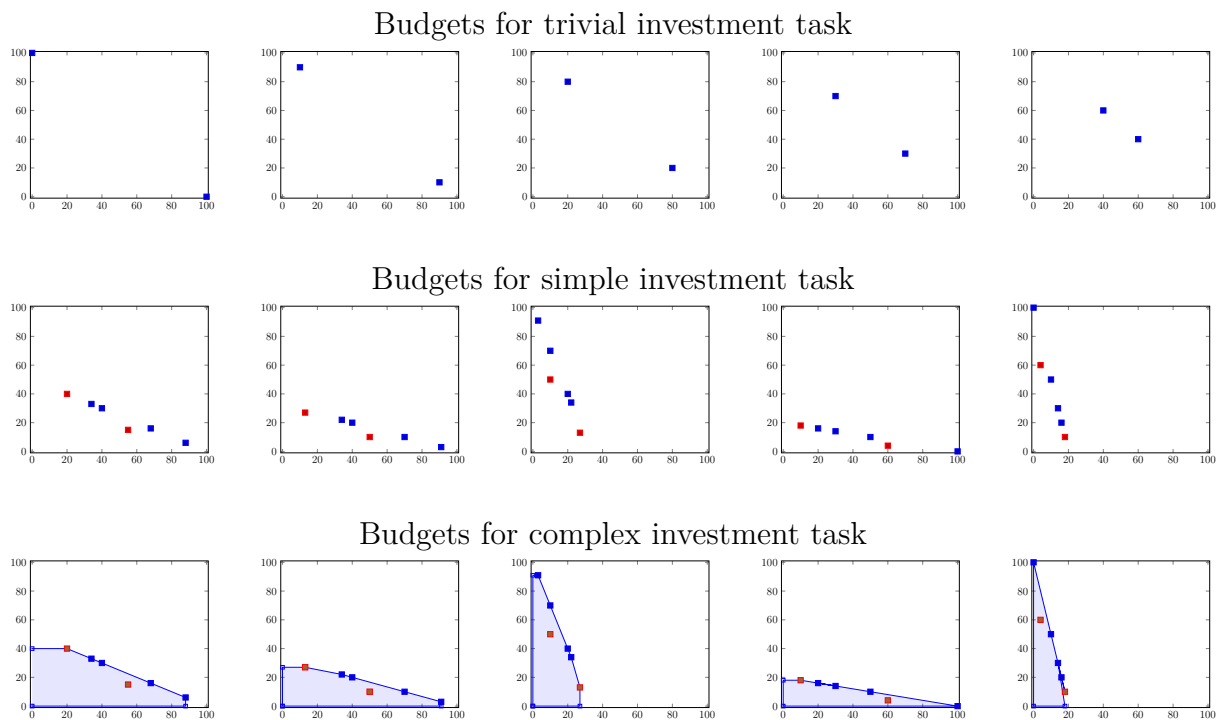
\begin{figure}[htbp]
    \centering
\begin{tabular}{ccccc}

\multicolumn{5}{c}{Budgets for trivial investment task}
\\
\resizebox{!}{.15\linewidth}{
    \begin{tikzpicture} \begin{axis}[
  xmin=-1, xmax=101,
  ymin=-1, ymax=101,
]  \addplot+[
  only marks,
  mark = square*,
  fill=blue,
  color=blue,
  mark size=2.9pt
  ]
  coordinates{
  (100,0)
  (0,100)};
  \end{axis}
  \end{tikzpicture}
}

&

\resizebox{!}{.15\linewidth}{
    \begin{tikzpicture} \begin{axis}[
  xmin=-1, xmax=101,
  ymin=-1, ymax=101,
]
\addplot+[
  only marks,
  mark = square*,
  fill=blue,
  color=blue,
  mark size=2.9pt
  ]
  coordinates{
  (90,10)
  (10,90)};
  \end{axis}
  \end{tikzpicture}
}

&

\resizebox{!}{.15\linewidth}{
    \begin{tikzpicture} \begin{axis}[
  xmin=-1, xmax=101,
  ymin=-1, ymax=101,
]  \addplot+[
  only marks,
  mark = square*,
  fill=blue,
  color=blue,
  mark size=2.9pt
  ]
  coordinates{
  (80,20)
  (20,80)};
  \end{axis}
  \end{tikzpicture}
}

&

\resizebox{!}{.15\linewidth}{
\begin{tikzpicture} \begin{axis}[
  xmin=-1, xmax=101,
  ymin=-1, ymax=101,
]
\addplot+[
  only marks,
  mark = square*,
  fill=blue,
  color=blue,
  mark size=2.9pt
  ]
  coordinates{
  (70,30)
  (30,70)};
  \end{axis}
  \end{tikzpicture}
}

&

\resizebox{!}{.15\linewidth}{
\begin{tikzpicture} \begin{axis}[
  xmin=-1, xmax=101,
  ymin=-1, ymax=101,
]
\addplot+[
  only marks,
  mark = square*,
  fill=blue,
  color=blue,
  mark size=2.9pt
  ]
  coordinates{
  (60,40)
  (40,60)};
  \end{axis}
  \end{tikzpicture}
}

\\
\multicolumn{5}{c}{}

\\
\multicolumn{5}{c}{Budgets for simple investment task}
\\

\resizebox{!}{.15\linewidth}{
\begin{tikzpicture} \begin{axis}[
  xmin=-1, xmax=101,
  ymin=-1, ymax=101,
]  \addplot+[
  only marks,
  mark = square*,
  fill=blue,
  color=blue,
  mark size=2.9pt
  ]
  coordinates{ 
  (68, 16)
  (40, 30)
  (88, 6)
  (34, 33)};
\addplot+[
  only marks,
  mark = square*,
  color=red,
  fill=red,
  mark size=2.9pt
  ]
  coordinates{
  (55, 15)
  (20, 40)
};
  \end{axis}
  \end{tikzpicture}
}
&

\resizebox{!}{.15\linewidth}{
\begin{tikzpicture} \begin{axis}[
  xmin=-1, xmax=101,
  ymin=-1, ymax=101,
]
\addplot+[
  only marks,
  mark = square*,
  fill=blue,
  color=blue,
  mark size=2.9pt
  ]
  coordinates{
  (40, 20)
  (91, 3)
  (34, 22)
  (70, 10)
  };
\addplot+[
  only marks,
  mark = square*,
  color=red,
  fill=red,
  mark size=2.9pt
  ]
  coordinates{
  (50, 10)
  (13, 27)
};
  \end{axis}
  \end{tikzpicture}
}

&

\resizebox{!}{.15\linewidth}{
\begin{tikzpicture} \begin{axis}[
  xmin=-1, xmax=101,
  ymin=-1, ymax=101,
]  \addplot+[
  only marks,
  mark = square*,
  fill=blue,
  color=blue,
  mark size=2.9pt
  ]
  coordinates{
  (10, 70)  
  (20, 40)
  (22, 34)
  (3, 91)};
\addplot+[
  only marks,
  mark = square*,
  color=red,
  fill=red,
  mark size=2.9pt
  ]
  coordinates{
  (10, 50)
  (27, 13)
};
  \end{axis}
  \end{tikzpicture}
}
&

\resizebox{!}{.15\linewidth}{
\begin{tikzpicture} \begin{axis}[
  xmin=-1, xmax=101,
  ymin=-1, ymax=101,
]
\addplot+[
  only marks,
  mark = square*,
  fill=blue,
  color=blue,
  mark size=2.9pt
  ]
  coordinates{
  (100, 0)
  (50, 10)
  (20, 16)
  (30, 14)
};
\addplot+[
  only marks,
  mark = square*,
  color=red,
  fill=red,
  mark size=2.9pt
  ]
  coordinates{
  (60, 4)
  (10, 18)
};
  \end{axis}
  \end{tikzpicture}
}

&

\resizebox{!}{.15\linewidth}{
\begin{tikzpicture} \begin{axis}[
  xmin=-1, xmax=101,
  ymin=-1, ymax=101,
]
\addplot+[
  only marks,
  mark = square*,
  fill=blue,
  color=blue,
  mark size=2.9pt
  ]
  coordinates{
  (16, 20)
  (0, 100)
  (10, 50)
  (14, 30)
  };
\addplot+[
  only marks,
  mark = square*,
  color=red,
  fill=red,
  mark size=2.9pt
  ]
  coordinates{
  (4, 60)
  (18, 10)
};
  \end{axis}
  \end{tikzpicture}
}

\\
\multicolumn{5}{c}{}

\\
\multicolumn{5}{c}{Budgets for complex investment task}
\\

\resizebox{!}{.15\linewidth}{
\begin{tikzpicture} \begin{axis}[
  xmin=-1, xmax=101,
  ymin=-1, ymax=101,
]  \addplot+[
  only marks,
  mark = square*,
  fill=blue,
  color=blue,
  mark size=2.9pt
  ]
  coordinates{
  (68, 16)
  (40, 30)
  (88, 6)
  (34, 33)};

\addplot+[
  fill=blue!60,
  color=blue,
  fill opacity=0.1
  ]
  coordinates{
  (0,0)
  (88,0)
  (88, 6)
  (68, 16)
  (40, 30)
  (34, 33)
  (20, 40)
  (0,40)
  };
\addplot+[
  only marks,
  mark = square*,
  color=red,
  fill=red,
  mark size=2.9pt
  ]
  coordinates{
  (55, 15)
  (20, 40)
};
  \end{axis}
  \end{tikzpicture}
}

&

\resizebox{!}{.15\linewidth}{
\begin{tikzpicture} \begin{axis}[
  xmin=-1, xmax=101,
  ymin=-1, ymax=101,
]
\addplot+[
  only marks,
  mark = square*,
  fill=blue,
  color=blue,
  mark size=2.9pt
  ]
  coordinates{
  (40, 20)
  (91, 3)
  (34, 22)
  (70, 10)
};

\addplot+[
  fill=blue!60,
  color=blue,
  fill opacity=0.1
  ]
  coordinates{
  (0,0)
  (91,0)
  (91, 3)
  (70, 10)
  (40, 20)
  (34, 22)
  (13, 27)
  (0,27)
  (0,0)
  };
\addplot+[
  only marks,
  mark = square*,
  color=red,
  fill=red,
  mark size=3pt
  ]
  coordinates{
  (50, 10)
  (13, 27)
};
  \end{axis}
  \end{tikzpicture}
}

&

\resizebox{!}{.15\linewidth}{
\begin{tikzpicture} \begin{axis}[
  xmin=-1, xmax=101,
  ymin=-1, ymax=101,
]  \addplot+[
  only marks,
  mark = square*,
  fill=blue,
  color=blue,
  mark size=2.9pt
  ]
  coordinates{
  (10, 70)
  (20, 40)
  (22, 34)
  (3, 91)};
\addplot+[
  fill=blue!60,
  color=blue,
  fill opacity=0.1
  ]
  coordinates{
  (0,0)
  (0,91)
  (3, 91)
  (10, 70)
  (20, 40)
  (22, 34)
  (27, 13)
  (27,0)
  (0,0)
  };
\addplot+[
  only marks,
  mark = square*,
  color=red,
  fill=red,
  mark size=2.9pt
  ]
  coordinates{
  (10, 50)
  (27, 13)
};
  \end{axis}
  \end{tikzpicture}
}

&

\resizebox{!}{.15\linewidth}{
\begin{tikzpicture} \begin{axis}[
  xmin=-1, xmax=101,
  ymin=-1, ymax=101,
]
\addplot+[
  only marks,
  mark = square*,
  fill=blue,
  color=blue,
  mark size=2.9pt
  ]
  coordinates{
  (100, 0)
  (50, 10)
  (20, 16)
  (30, 14)
};
\addplot+[
  fill=blue!60,
  color=blue,
  fill opacity=0.1
  ]
  coordinates{
  (100, 0)
  (50, 10)
  (20, 16)
  (30, 14)
  (10, 18)
  (0,18)
  (0,0)
  (100,0)
  };
\addplot+[
  only marks,
  mark = square*,
  color=red,
  fill=red,
  mark size=2.9pt
  ]
  coordinates{
  (60, 4)
  (10, 18)
};
  \end{axis}
  \end{tikzpicture}
}
&

\resizebox{!}{.15\linewidth}{
\begin{tikzpicture} \begin{axis}[
  xmin=-1, xmax=101,
  ymin=-1, ymax=101,
]
\addplot+[
  only marks,
  mark = square*,
  fill=blue,
  color=blue,
  mark size=2.9pt
  ]
  coordinates{
  (16, 20)
  (0, 100)
  (10, 50)
  (14, 30)
  
  };
\addplot+[
  fill=blue!60,
  color=blue,
  fill opacity=0.1
  ]
  coordinates{
  (0, 100)
  (10, 50)
  (18, 10)
  (16, 20)
  (14, 30)
  (18,0)
  (0,0)
  (0,100)
  };
\addplot+[
  only marks,
  mark = square*,
  color=red,
  fill=red,
  mark size=2.9pt
  ]
  coordinates{
  (4, 60)
  (18, 10)
};
  \end{axis}
  \end{tikzpicture}
}

\end{tabular}
\caption{Budget specifications used in the experiment}
\label{fig:BudgetsForExperiment}
\end{figure}